\documentclass[pre,aps,twocolumn,superscriptaddress]{revtex4}

\usepackage[dvips]{graphicx}
\usepackage{amssymb,amsfonts,amsmath}
\usepackage{color}
\usepackage{ulem}
\usepackage[hidelinks]{hyperref}
\usepackage{bbm}
\usepackage{dsfont}
\usepackage{bm}
\usepackage{xcolor}

\newcommand{\rv}{{\mathbf r}}

\newcommand{\ev}{{\bf e}}

\newcommand{\Tr}{{\rm Tr}\,}
\newcommand{\e}{{\rm e}}

\newcommand{\pv}{{\bf p}}

\newcommand{\Fv}{{\bf F}}

\newcommand{\fv}{{\bf f}}

\newcommand{\eps}{{\boldsymbol \epsilon}}
\newcommand{\unity}{{\mathds 1}}
\newcommand{\cov}{{\rm cov}}
\newcommand{\Vext}{V_{\rm ext}}
\newcommand{\ext}{{\rm ext}}

\newcommand{\avg}[1]{\Big\langle #1 \Big\rangle}
\newcommand{\eqr}[1]{Eq.~\eqref{#1}}
\newcommand{\ff}{{f\!f}}
\newcommand{\gradf}{{\nabla\!f}}
\newcommand{\gff}{{\sf g}_\ff}
\newcommand{\ggradf}{{\sf g}_\gradf}

\newcommand{\mydelete}[1]{{}}
\newcommand{\taub}{{\boldsymbol\tau}}

\newcommand{\rmint}{{\rm int}}
\newcommand{\rmext}{{\rm ext}}
\newcommand{\rmself}{{\rm self}}

\begin{document}

\title{Noether invariance theory for the equilibrium force structure of soft matter}

\author{Sophie Hermann}
\affiliation{Theoretische Physik II, Physikalisches Institut, 
  Universit{\"a}t Bayreuth, D-95447 Bayreuth, Germany}
\author{Florian Samm\"uller}
\affiliation{Theoretische Physik II, Physikalisches Institut, 
  Universit{\"a}t Bayreuth, D-95447 Bayreuth, Germany}
\author{Matthias Schmidt}
\affiliation{Theoretische Physik II, Physikalisches Institut, 
  Universit{\"a}t Bayreuth, D-95447 Bayreuth, Germany}
\email{Matthias.Schmidt@uni-bayreuth.de}

\date{26 January 2024, revised version: 23 March 2024}

\begin{abstract}
We give details and derivations for the Noether invariance theory that
characterizes the spatial equilibrium structure of inhomogeneous
classical many-body systems, as recently proposed and investigated for
bulk systems
[\href{https://doi.org/10.1103/PhysRevLett.130.268203}{F. Samm\"uller
    {\it et al.}, Phys. Rev. Lett. {\bf 130}, 268203 (2023)}]. Thereby
an intrinsic thermal symmetry against a local shifting transformation
on phase space is exploited on the basis of the Noether theorem for
invariant variations.  We consider the consequences of the shifting
that emerge at second order in the displacement field that
parameterizes the transformation.  In a natural way the standard
two-body density distribution is generated. Its second spatial
derivative (Hessian) is thereby balanced by two further and different
two-body correlation functions, which respectively introduce 
thermally averaged force correlations and force gradients in a
systematic and microscopically sharp way into the framework.  Separate
exact self and distinct sum rules hold expressing this balance.  We
exemplify the validity of the theory on the basis of computer
simulations for the Lennard-Jones gas, liquid, and crystal, the
Weeks-Chandler-Andersen fluid, monatomic Molinero-Moore water at
ambient conditions, a three-body-interacting colloidal gel former, the
Yukawa and soft-sphere dipolar fluids, and for isotropic and nematic
phases of Gay-Berne particles.  We describe explicitly the derivation
of the sum rules based on Noether's theorem and also give more
elementary proofs based on partial phase space integration following
Yvon's theorem.
\end{abstract}

\maketitle

\section{Introduction}

The spatial structure of soft matter can be very rich and varied. For
a fluid at the most fundamental level the bulk structure is described
by the pair (or radial) distribution function $g(r)$, which is a
measure of the probability of finding two particles separated by a
distance~$r$~\cite{hansen2013, frenkel2023book}. The information
contained in $g(r)$ is analogously contained in the static structure
factor $S(k)$, which is directly connected to the Fourier transform of
$g(r)$ and is accessible in scattering experiments. Real-space
measurements of pair correlation functions in colloidal systems give
much insight into the mesoscopic particle structure
\cite{royall2007gofr, thorneywork2014, statt2016}.

The behaviour of $g(r)$ can be systematically analyzed and classified
by asymptotic expansion at large interparticle distances, which allows
to identify different types of decay \cite{evans1993decay,
  evans1994decay, dijkstra2000decay, grodon2004decay,
  cats2021decay}. The asymptotic analysis is fundamental to the
behaviour of the system at hand, as the results remain relevant for
the decay of the one-body density profile into the bulk far away from
a substrate or solute that created the density inhomogeneity.
Furthermore $g(r)$ plays the prominent role to determine the
thermodynamics and equation of state via spatial integration according
to the virial or compressibility routes to the pressure as are
elementary ingredients of liquid integral equation
theory~\cite{hansen2013}.

Despite these virtues, there is much activity of going beyond $g(r)$
in the spatial characterization of soft matter.  Recent efforts
include characterizing the structure of liquids using four-point
correlation functions \cite{zhang2020pnas}. Such measures were argued
to be relevant for the dynamics in dense colloidal liquids
\cite{singh2023pnas}.  The packing efficiency of binary hard sphere
systems was related to fluid structure at intermediate ranges
\cite{yuan2021prl}. Emergent structural correlations in dense liquids
were characterized up to the six-body structure factor
\cite{pihlajamaa2023}. Using machine learning the averaged structural
features centered around nearby particles were used to predict
dynamics in supercooled liquids \cite{boattini2011}.

Noether's theorem of invariant variations \cite{noether1918,
  byers1998} was put in a statistical physics setting in a number of
different ways \cite{baez2013markov, marvian2014quantum, sasa2016,
  sasa2019, revzen1970, baez2020bottom, bravetti2023}. The recent
thermal Noether invariance theory \cite{hermann2021noether,
  hermann2022topicalReview, hermann2022variance, hermann2022quantum,
  tschopp2022forceDFT, sammueller2023whatIsLiquid, robitschko2024any}
is based on the identification of the symmetry of the statistical
many-body systems against specific shifting and rotation operations on
phase space. Corresponding force and torque correlation functions are
generated and exact statistical mechanical identities (``sum rules'')
play the role that in more conventional uses of the Noether theorem
are played by the resulting conservation laws.  The recent hyperforce
approach by Robitschko {et al.}~\cite{robitschko2024any} carries the
thermal Noether concept further in that it allows to address the
behaviour of arbitrary observables in equilibribum.

\begin{figure*}[htb!]
      \includegraphics[height=11.5cm]{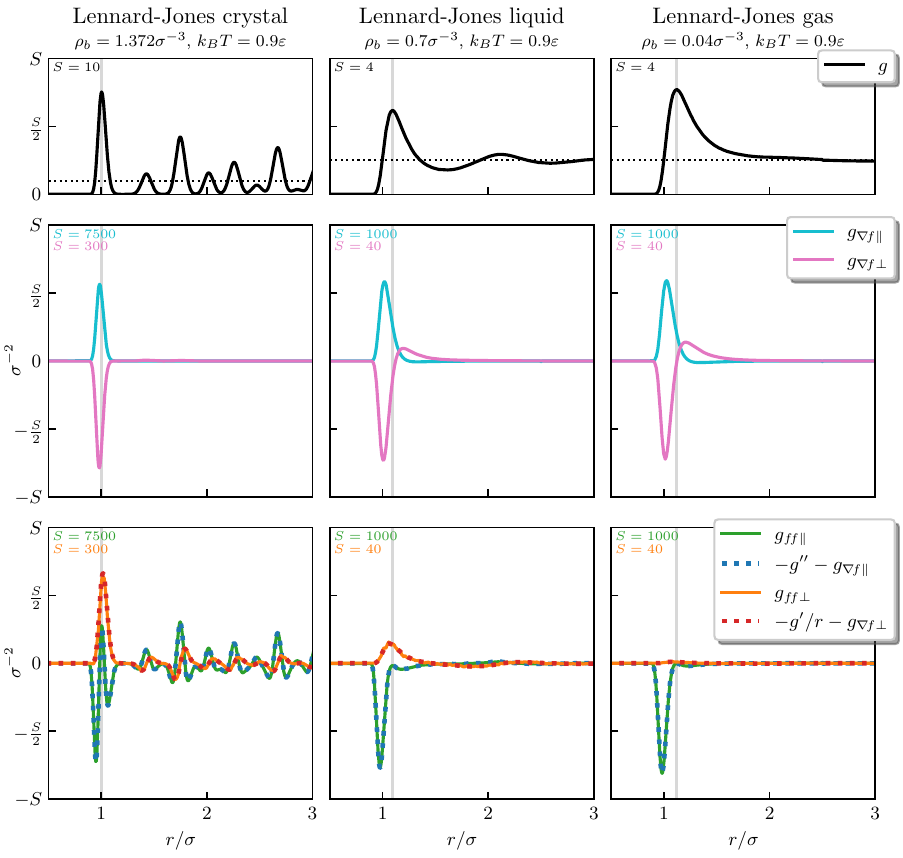}
\caption{Illustrative examples of the Noether force correlation
  functions for the Lennard-Jones system in the fcc crystal phase
  (first colunm), the liquid phase (second column), and the gas phase
  (third column).  Shown are results for the standard pair correlation
  function $g(r)$ in all three phases (first row). The Noether
  invariance theory introduces two further pair correlation functions,
  ${\sf g}_\gradf(r)$ and ${\sf g}_\ff(r)$, as are defined in detail
  in Sec.~\ref{SECbulkLiquidNoetherStructure}. These tensorial
  functions possess radial ($\parallel$) and transversal ($\perp$)
  components, both of which are plotted.  Briefly, the two-body
  force-gradient correlator ${\sf g}_\gradf(r)$ (second row) is a
  measure of the change of the force on one particle upon changing the
  position of a second particle that is located at a distance $r$ from
  the first particle. The force-force pair correlator ${\sf g}_\ff(r)$
  (third row) associates the individual forces in the particle pair
  with each other.  That the dotted lines (third panel) coincide with
  the full lines verifies the 3g-rule \eqref{EQ3gRuleIntroduction}. In
  the figure the respective vertical scale factor $S$ is given in the
  top left corner of each panel and the scaled values for bulk density
  $\rho_b$ and temperature $T$ are indicated for each statepoint above
  the respective column. The vertical gray lines indicate the position
  of the first maximum of $g(r)$ as a guide to the eye.  Adapted from
  the Supplementary Material of
  Ref.~\cite{sammueller2023whatIsLiquid}.
   \label{FIG1}
  }
\end{figure*}

The classical question ``What is liquid?''~\cite{barker1976} appears
in new light when regarded from the viewpoint of the thermal Noether
invariance theory. In particular the recently formulated force-force
and force-gradient pair correlation functions
\cite{sammueller2023whatIsLiquid} give fresh insight into the
correlation structure. These correlation functions are generated from
canonical ``shifting'' transformation on phase space
\cite{goldstein2002} and thereby considering the second order in the
spatial shifting field. The shifting field is an important formal
device for formulating and exploiting the Noether invariance, yet the
resulting identities are free of the shifting field. Only genuine
system properties are correlated with each other and the framework is
entirely mechanical.  In particular forces are at the forefront, as
these appear naturally in response to spatial displacement (shifting)
of the free energy. Forces comprise interparticle, external, and
diffusive contributions and they occur both globally as well as
locally resolved in arbitrary spatially inhomogeneous systems.

For bulk systems the second-order Noether invariance reduces to the
``3g-rule'' that relates three different types of pair functions with
each other (we reproduce from
Sec.~\ref{SECbulkLiquidNoetherStructure}):
\begin{align}
  \nabla\nabla g(r) & + {\sf g}_\gradf(r) + {\sf g}_\ff(r) = 0.
  \label{EQ3gRuleIntroduction}
\end{align}
Here $\nabla$ indicates the spatial gradient, such that $\nabla\nabla
g(r)$ is the Hessian of the pair correlation function. Furthermore the
force-gradient pair correlator ${\sf g}_\gradf(r)$ and the force-force
pair correlator ${\sf g}_\ff(r)$ are both $3\times 3$-matrices. These
correlation functions also address particle pairs, but they
incorporate additional information over the mere counting in
$g(r)$. In particular $\gff(r)$ correlates the interparticle forces
that act on the two particles in the pair with each other. The
force-gradient correlation function $\ggradf(r)$ is the mean gradient
of the force on one of the particles with respect to changing the
position of the partner. A more detailed description is given below.

The Noether 3g-rule \eqref{EQ3gRuleIntroduction} was verified by
Samm\"uller {et al.}~\cite{sammueller2023whatIsLiquid} across a wide
range of model Hamiltonians, including Lennard-Jones, Yukawa,
soft-sphere dipolar, Stockmayer, Gay-Berne, and
Weeks-Chandler-Andersen liquids. Also Hamiltonians which incorporate
three-body interaction terms were considered, such as the monatomic
Molinero-Moore water \cite{molinero2009, coe2022water} and a colloidal
gel forming model proposed by Kob and coworkers \cite{saw2009,
  saw2011, sammueller2023gel}. The identity
\eqref{EQ3gRuleIntroduction} holds beyond fluids also in liquid
crystals and in solids.  Besides certain very generic features, such
as a depletion at small separation, the individual models and phases
display differences in shapes and forms of $g(r)$, as expected. The
observed differences in $\gff(r)$ and in $\ggradf(r)$ across different
models and phases are significantly more pronounced though and they
are indicative of a wide range of specific features, including the
presence of interparticle attraction and of particle strand formation
in colloidal gels \cite{jadrich2023preface} and in dipolar fluids
\cite{klapp2005topRev, stevens1995, teixeira2000topRev, tavares1999,
  allen2019topRev}.

To illustrate this behaviour we reproduce in Figure~\ref{FIG1} results
of Ref.~\cite{sammueller2023whatIsLiquid} for the Lennard-Jones system
in all of its three stable thermodynamic phases: crystal, liquid, and
gas. The pair potential is thereby truncated and shifted at a
  cutoff distance of $r_c=2.5\sigma$, where~$\sigma$ denotes the
  particle size. The data is obtained from equilibribum sampling via
adaptive Brownian dynamics \cite{sammueller2021}, which allows for
tight error control of the occurring forces. The behaviour of $g(r)$
in the gas can be well-understood from the virial expansion where to
lowest order $g(r)=\exp(-\beta\phi(r))$, where $\beta$ denotes inverse
temperature and $\phi(r)$ is the (Lennard-Jones) pair potential.

The force-force and force-gradient pair correlation functions feature
tensorial components that are radial and transversal with respect to
the difference vector between both particle centers.  
 Due to the rotational symmetry of fluid states and the absence of
  any chirality, these are the only two relevant tensor components to
  consider. In principle the structure of the crystal is much richer,
  but we project onto the same two tensor components and hence average
  over the further spatial and rotational inhomogeneities of the
  crystal.
The mean force gradient $\ggradf(r)$ vanishes beyond the range of the
(truncated) Lennard-Jones potential (second row in Fig.~\ref{FIG1})
and it features a clear sign of the interparticle attraction in the
transversal component; see the positive peak of the pink curve. The
force-force pair correlation function $\gff(r)$ (third row in
Fig.~\ref{FIG1}) extends beyond the range of truncation and it again
shows very rich correlation structure. In the symmetry-broken state of
the fcc crystal the results are obtained from only resolving according
to scalar distances. While this implies loss of information over the
full inhomogeneous two-body resolution, the results shown in
Fig.~\ref{FIG1} for the Lennard-Jones crystal nevertheless indicate
intricate and long-ranged force-force correlation behaviour. In
striking contrast, the force-gradient correlation function ${\sf
  g}_\gradf(r)$ remains short-ranged and it strictly vanishes beyond
the cutoff for short-ranged or truncated pair potentials.
 The accordance of the dashed and the solid lines in the third row
  of Fig.~\ref{FIG1} confirms numerically that the 3g-rule sum
  rule~\eqref{EQ3gRuleIntroduction} is satisfied.

With the central mathematical ideas only being sketched in
Ref.~\cite{sammueller2023whatIsLiquid}, we here give a detailed
account of the second-order thermal Noether invariance theory. We
present explicit derivations as well as several new results, such as
the density-force correlation sum rule
\eqref{EQdensityForceCorrelationSumRule}.  We also give proofs of the
Noether identities from partial integration on phase space in analogy
to the Yvon theorem~\cite{hansen2013}. This elementary route is
conceptually simple, as it only requires direct manipulation of the
phase space integrals, and it hence provides a valuable consistency
check. We also describe in detail the simulation results for the
variety of models considered.

The paper is organized as follows.  In Sec.~\ref{SECtheoreticalSetup}
we lay out the statistical many-body physics under investigation and
describe the theoretical setup of thermal Noether invariance against
local shifting.  The two-body sum rules that follow from the
second-order invariance are presented in
Sec.~\ref{SEClocalSecondOrderInvariance}.  A general density-force
correlation sum rule is proven in Sec.~\ref{SECdensityForceSumRule}.
The emerging force correlators are split into separate contributions
in Sec.~\ref{SECforceCorrelatorSplitting}, which facilitates to prove
the general Noether two-body sum rules presented in
  Sec.~\ref{SEClocalSecondOrderInvariance} from first principles.
The general inhomogeneous two-body framework is specialized to the
case of bulk systems in Sec.~\ref{SECbulkLiquidNoetherStructure} and
we present and discuss our simulation results therein.  Several
important yet technical points are shown in five Appendices. This
includes considerations of the kinetic stress autocorrelations
presented in Appendix~\ref{SECkineticEnergySingularity}. Kinetic
energy curvature terms are addressed in
Appendix~\ref{SECkineticSelfCurvature} and potential energy curvature
contributions are presented in
Appendix~\ref{SECpotentialEnergyCurvature}.  Alternative explicit
proofs of the Noether sum rules based on partial integration are given
in Appendix~\ref{SECviaPartialIntegration}.  Details about measuring
correlation functions in simulations are given in
Appendix~\ref{SECappendixMeasuring}. We present our conclusions in
Sec.~\ref{SECconclusions}.

\section{Theoretical setup}
\label{SECtheoreticalSetup}

We consider systems of $N$ particles with position coordinates
$\rv_1,\ldots,\rv_N\equiv \rv^N$ and linear momenta
$\pv_1,\ldots,\pv_N\equiv \pv^N$. The Hamiltonian $H$ is taken to
contain kinetic energy, as well as interparticle and external energy
contributions,
\begin{align}
  H &= \sum_i \frac{\pv_i^2}{2m} + u(\rv^N) 
  + \sum_i V_{\rm ext}(\rv_i),
  \label{EQHamiltonian2}
\end{align}
where the summation index $i=1,\ldots,N$ ranges over all particles,
$m$~denotes the particle mass, $u(\rv^N)$ is the interparticle
interaction potential, and the one-body external potential $V_{\rm
  ext}(\rv)$ is given as a function of a single (generic) position
variable~$\rv$.

The canonical transformation \cite{goldstein2002} of the phase space
coordinates that was put forward in Refs.~\cite{hermann2022quantum,
  tschopp2022forceDFT, sammueller2023whatIsLiquid} maps original onto
new coordinates as well as original onto new momenta according to the
following map:
\begin{align}
  \rv_i &\to \rv_i + \eps(\rv_i) = \tilde\rv_i,
  \label{EQcanonicalTransformationCoordinates}\\
  \pv_i &\to [\unity + \nabla_i\eps(\rv_i)]^{-1}\cdot\pv_i = \tilde\pv_i,
  \label{EQcanonicalTransformationMomenta}
\end{align}
where the tilde indicates the new variables. The three-dimensional
vector field $\eps(\rv)$ parameterizes the transform, $\unity$ denotes
the $3\times 3$-unit matrix, $\nabla_i$ indicates the derivative with
respect to $\rv_i$, and matrix inversion is indicated by the
superscript $-1$. Here we consider systems in three space dimensions
but the theory is general. The transformation
\eqref{EQcanonicalTransformationCoordinates} and
\eqref{EQcanonicalTransformationMomenta} preserves the phase space
volume element as well as thermal averages~\cite{tschopp2022forceDFT,
  hermann2022quantum, sammueller2023whatIsLiquid, goldstein2002} and
it constitutes a {\it canonical transformation} in the sense of
classical mechanics \cite{goldstein2002}. This property can e.g.\ be
seen explicitly by considering a generator of the form ${\cal G} =
\sum_{i=1}^N \tilde\pv_i\cdot (\rv_i +\eps(\rv_i))$, where as before
the tilde indicates the new phase space variables.  Applying the
generic transformation relations $\tilde\rv_i = \partial {\cal
  G}/\partial \tilde\pv_i$ and $\pv_i = \partial {\cal G}/\partial
\rv_i$ \cite{goldstein2002} and solving for the new variables then
yields the transformation equations
\eqref{EQcanonicalTransformationCoordinates} and
\eqref{EQcanonicalTransformationMomenta}. The form of the
  shifting vector field $\eps(\rv)$ must thereby such that the
  transformation is bijective \cite{hermann2022quantum}.

We Taylor expand the transformation in the shifting field
$\eps(\rv)$. The coordinate transformation
\eqref{EQcanonicalTransformationCoordinates} is already linear in
$\eps(\rv)$ and it is hence unaffected.  The momentum
transformation~\eqref{EQcanonicalTransformationMomenta} can be
expanded as a Neumann series, which is the analog of the geometric
series for matrices or, more generally, for bounded operators.
Assuming that both the shifting field and its gradient are small, up
to second order the expansion yields:
\begin{align}
  [\unity + \nabla_i\eps(\rv_i)]^{-1}
  =\unity - \nabla_i\eps(\rv_i)+
  [\nabla_i\eps(\rv_i)]^2-\ldots,
  \label{EQneumannSeries}
\end{align}
where the square on the right hand side denotes matrix
(self-)multiplication, i.e., $[\nabla_i\eps(\rv_i)]^2 =
[\nabla_i\eps(\rv_i)] \cdot [\nabla_i\eps(\rv_i)]$. Verifying
\eqr{EQneumannSeries} to second order in the shifting gradient is
straightforward via matrix multiplying its both sides by $\unity +
\nabla_i \eps(\rv_i)$, which per definition of the inverse gives
$\unity$ on the left hand side. The right hand side gives the same
result up to second order upon carrying out the matrix multiplications
and simplifying.

When expressed in the new variables, the Hamiltonian becomes
functionally dependent on the shifting field, $H\to H[\eps]$.  We
first consider the term linear in $\eps(\rv)$
\cite{hermann2022quantum, tschopp2022forceDFT,
  sammueller2023whatIsLiquid}. One sees that the position-resolved
one-body force density operator $\hat\Fv(\rv)$ follows from functional
differentiation according to:
\begin{align}
  -\frac{\delta H[\eps]}{\delta \eps(\rv)}  \Big|_{\eps=0}
  &= 
  \hat\Fv(\rv),
  \label{EQFoperatorFromDifferentiation}
\end{align}
where $\delta/\delta\eps(\rv)$ indicates the functional derivative
with respect to the shifting field $\eps(\rv)$; see
e.g.\ Ref.~\cite{schmidt2022rmp} for an overview of functional
differentiation techniques. After the derivative is taken, $\eps(\rv)$
is set to zero, as indicated in the notation on the left hand side of
\eqr{EQFoperatorFromDifferentiation}. The structure of the force
density operator $\hat\Fv(\rv)$ mirrors that of the Hamiltonian
\eqref{EQHamiltonian2} in that it contains kinetic, interparticle, and
external contributions:
\begin{align}
  \hat\Fv(\rv) &=
  \nabla\cdot\hat\taub(\rv)
  + \hat\Fv_{\rm int}(\rv)
  -\hat\rho(\rv)\nabla V_{\rm ext}(\rv),
  \label{EQForceDensityOperator}
\end{align}
where $\nabla$ denotes the derivative with respect to $\rv$.  The
one-body operators for the kinetic stress $\hat\taub(\rv)$, the
interparticle force density $\hat\Fv_{\rm int}(\rv)$
\cite{schmidt2022rmp}, and the microscopic density distribution
$\hat\rho(\rv)$ \cite{evans1979,hansen2013} are given respectively by
\begin{align}
  \hat \taub(\rv) &= -\sum_i \frac{\pv_i\pv_i}{m}\delta(\rv-\rv_i),
  \label{EQkineticStressOperator}
  \\
  \hat\Fv_{\rm int}(\rv) &= 
  -\sum_i\delta(\rv-\rv_i)\nabla_iu(\rv^N),
  \label{EQFintOperator}
  \\
  \hat\rho(\rv) &= \sum_i\delta(\rv-\rv_i),
  \label{DEdensityOperator}
\end{align}
where $\delta(\cdot)$ indicates the (three-dimensional) Dirac
distribution and $\pv_i\pv_i$ is the momentum of particle $i$
dyadically multiplied with itself.

The following analysis within Statistical Mechanics holds equally well
with fixed number of particles, i.e., in the canonical ensemble. To be
explicit, we here formulate our theory in the grand ensemble at
temperature~$T$ and chemical potential $\mu$. The grand potential
$\Omega$ and the grand partition sum $\Xi$ are given respectively by
\begin{align}
  \Omega &= -k_BT\ln\Xi,
  \label{EQgrandPotential}\\
  \Xi &= \Tr {\rm e}^{-\beta(H-\mu N)},
  \label{EQgrandPartitionSum}
\end{align}
where $k_B$ is the Boltzmann constant, $\beta=1/(k_BT)$ denotes
inverse temperature, and the classical ``trace'' operation in the
grand ensemble is given by $\Tr \cdot=\sum_{N=0}^\infty (N!
h^{3N})^{-1}\int d\rv_1\ldots d \rv_N \int d\pv_1\ldots d\pv_N \cdot$,
where $h$ indicates the Planck constant. The corresponding grand
probability distribution (Gibbs measure) is $\Psi={\rm
  e}^{-\beta(H-\mu N)}/\Xi$. Thermal averages are defined via
$\langle\cdot\rangle=\Tr \Psi \cdot$, such that the density profile is
the average of the one-body density operator and written as
$\rho(\rv)=\langle\hat\rho(\rv)\rangle$.  Correspondingly the average
kinetic stress is $\taub(\rv)=\langle\hat\taub(\rv)\rangle$ and the
mean interparticle force density is $\Fv_{\rm
  int}(\rv)=\langle\hat\Fv_{\rm int}(\rv)\rangle$.

At face value the grand partition sum \eqref{EQgrandPartitionSum}
carries an apparent functional dependence on the shifting field
$\eps(\rv)$, via the occurrence of $\eps(\rv)$ in the transformed
Hamiltonian~$H[\eps]$ \cite{hermann2022quantum, tschopp2022forceDFT,
  sammueller2023whatIsLiquid}. We hence also have a functional
dependence of the grand partition sum, $\Xi[\eps]$, and consequently
of the grand potential, $\Omega[\eps]$, as it is generated via the
standard relationship~\eqref{EQgrandPotential}.  We functionally
Taylor expand the grand potential in the shifting field to quadratic
order, which gives
\begin{align}
  \Omega[\eps] &= \Omega 
  + \int d\rv
  \frac{\delta\Omega[\eps]}{\delta\eps(\rv)}\Big|_{\eps=0}
  \cdot\eps(\rv)
  \label{EQOmegaEpsExpansion}\\
  &\qquad+\frac{1}{2}\int d\rv d\rv'
  \frac{\delta^2\Omega[\eps]}{\delta\eps(\rv)\delta\eps(\rv')}
  \Big|_{\rm \eps=0}
  :\eps(\rv')\eps(\rv)+\ldots.\notag
\end{align}
The colon denotes a double tensor contraction and~$\Omega$, with no
argument, indicates the original grand
potential~\eqref{EQgrandPotential} with no transformation applied,
$\Omega=\Omega[0]$, where the functional argument $0$ indicates the
special case of vanishing shifting field, $\eps(\rv)=0$.

Despite the apparent dependence on $\eps(\rv)$, Noether invariance
\cite{hermann2021noether,hermann2022topicalReview} ascertains that the
grand potential remains unchanged upon applying the transformation,
and hence
\begin{align}
  \Omega[\eps] = \Omega,
  \label{EQOmegaInvariant}
\end{align}
which holds true for any (permissible) form of $\eps(\rv)$. As a
consequence, all terms in the functional expansion
\eqref{EQOmegaEpsExpansion} need to vanish identically. At first order
in the displacement field, we hence have
\begin{align}
  \frac{\delta\Omega[\eps]}{\delta\eps(\rv)} \Big|_{\eps=0} &= 0.
  \label{EQOmegaDerivativeZero}
\end{align}
Explicitly carrying out the functional derivative on the left hand of
\eqr{EQOmegaDerivativeZero} gives \cite{tschopp2022forceDFT,
  hermann2022quantum, sammueller2023whatIsLiquid}:
\begin{align}
  \frac{\delta\Omega[\eps]}{\delta\eps(\rv)} \Big|_{\eps=0} &=
  \Big\langle
  \frac{\delta H[\eps]}{\delta\eps(\rv)}\Big|_{\eps=0}
  \Big\rangle  =
  -\langle\hat\Fv(\rv)\rangle,
  \label{EQOmegaDerivativeLinear}
\end{align}
where we recall the total force density operator~$\hat\Fv(\rv)$ in the
form \eqref{EQForceDensityOperator}. The first equality in
Eq.~\eqref{EQOmegaDerivativeLinear} follows from applying the
definition \eqref{EQgrandPotential} of the grand potential. The second
equality is the thermal average of the total force density operator
identity \eqref{EQFoperatorFromDifferentiation}.

The expression on the right hand side of \eqr{EQOmegaDerivativeLinear}
constitutes the negative averaged one-body force density
$-\Fv(\rv)=-\langle\hat\Fv(\rv)\rangle$. Due to the invariance that is
expressed via \eqr{EQOmegaDerivativeZero} we can conclude that the
mean one-body force density vanishes in equilibribum,
\begin{align}
  \Fv(\rv) &= 0.
  \label{EQFisZero}
\end{align}
The explicit form \eqref{EQForceDensityOperator} of $\hat\Fv(\rv)$
allows to decompose the left hand side of \eqr{EQFisZero} into a sum
of ideal gas, interparticle, and external terms according to:
\begin{align}
  -k_BT\nabla\rho(\rv) + \Fv_{\rm int}(\rv) 
  - \rho(\rv)\nabla V_{\rm ext}(\rv)
  &= 0.
  \label{EQforceDensityBalance}
\end{align}
The force density balance \eqref{EQforceDensityBalance} is the analog
of the first member of the classical one-body Yvon-Born-Green equation
of liquid state theory~\cite{hansen2013}, in which $\Fv_\rmint(\rv)$
is further expressed in integral form.

The present functional route \cite{tschopp2022forceDFT,
  hermann2022quantum, sammueller2023whatIsLiquid} identifies the
equilibribum force density balance \eqref{EQforceDensityBalance} as
the first-order Noether invariance sum rule of the grand potential
with respect to the local shifting
\eqref{EQcanonicalTransformationCoordinates} and
\eqref{EQcanonicalTransformationMomenta} of phase space. This
transformation is canonical and hence the associated symmetry forms a
deeply rooted property of the statistical physics generated by the
standard Hamiltonian \eqref{EQHamiltonian2}.

As an aside, the recent force-based density functional theory
\cite{tschopp2022forceDFT, sammueller2022forceDFT} takes this concept
as the fundamental building block for making predictions based on an
excess (over ideal gas) free energy density functional.  Thereby the
interparticle force density is re-written according to the
Yvon-Born-Green hierarchy \cite{yvon1935, Born1946, hansen2013} as an
integral over the pair force times the two-body density, which in turn
is obtained from numerical solution of the inhomogeneous two-body
Ornstein-Zernike equation. Force-density functional theory has been
exemplified for inhomogeneous hard sphere systems in planar geometry
by inputting the two-body direct correlation functional from
fundamental measure theory \cite{rosenfeld1989, roth2010}, see
Refs.~\cite{tschopp2022forceDFT, sammueller2022forceDFT}.

\section{Local second-order invariance}
\label{SEClocalSecondOrderInvariance}

The previous Sec.~\ref{SECtheoreticalSetup} summarizes the theory
developed in Refs.~\cite{tschopp2022forceDFT, hermann2022quantum},
which we require for considering the second order. We recall that for
global shifting with spatially uniform displacement,
$\eps(\rv)=\eps_0=\rm const$, addressing the second order in $\eps_0$
yields a sum rule which relates the global force variance with the
mean global potential energy curvature \cite{hermann2022variance}.

In this global case the interparticle force contributions vanish and
only the external force field contributes, according to the following
global second-order sum rule:
\begin{align}
  \int d\rv d\rv' H_2(\rv,\rv')
  \nabla V_\ext(\rv) &\nabla' V_\ext(\rv') =
  \notag\\
  & k_BT \int d\rv \rho(\rv) \nabla\nabla V_\ext(\rv),
  \label{EQsumRuleGlobalSecondOrder}
\end{align}
where $\nabla'$ denotes the derivative with respect to $\rv'$ and
$H_2(\rv,\rv')=\cov(\hat\rho(\rv),\hat\rho(\rv'))$ is the standard
two-body correlation function of density fluctuations around the local
density profile~\cite{hansen2013, schmidt2022rmp}. Thereby the
covariance of two observables, as represented by phase space functions
$\hat A$ and $\hat B$, is defined in the standard way as $\cov(\hat
A,\hat B)=\langle \hat A\hat B\rangle- \langle\hat A\rangle \langle
\hat B\rangle$.  Upon carrying out the position integrals the left
  hand side of \eqr{EQsumRuleGlobalSecondOrder} can be alternatively
written as $\langle \hat \Fv_\rmext^0 \hat \Fv_\rmext^0 \rangle$,
where the global external force density operator is defined as
$\hat\Fv_\rmext^0 = -\sum_i \nabla_i V_\rmext(\rv_i)$ and the
covariance reduces to the mean of the product, because of the
vanishing mean external force in equilibribum, $\langle\hat
\Fv_\rmext^0\rangle=0$~\cite{hermann2021noether}.

Following the strategy of Ref.~\cite{sammueller2023whatIsLiquid} we
here describe in detail the consequences of locally resolved shifting
at second order in the displacement field. We recall that the
functional expansion~\eqref{EQOmegaEpsExpansion} of the grand
potential holds irrespective of the precise form of
$\eps(\rv)$. Addressing the second order term, this implies that its
prefactor in \eqr{EQOmegaEpsExpansion} needs to vanish identically:
\begin{align}
  \frac{\delta^2\Omega[\eps]}{\delta\eps(\rv)\delta\eps(\rv')}
  \Big|_{\eps=0} &= 0.
  \label{EQomegaSecondDerivative1}
\end{align}
Making the functional derivative that appears on the left hand side
more explicit yields
\begin{align}
  \frac{\delta^2\Omega[\eps]}{\delta\eps(\rv)\delta\eps(\rv')}  &=
  -\beta\,\cov\Big(
  \frac{\delta H[\eps]}{\delta\eps(\rv)},
  \frac{\delta H[\eps]}{\delta\eps(\rv')}
  \Big)+
  \Big\langle
  \frac{\delta^2 H[\eps]}{\delta\eps(\rv)\delta\eps(\rv')}
  \Big\rangle.
  \label{EQomegaSecondDerivative2}
\end{align}
The functional derivative of the Hamiltonian, $\delta
H[\eps]/\delta\eps(\rv)$, can be expressed as the negative force
density operator via \eqr{EQFoperatorFromDifferentiation}. Then
inserting \eqr{EQomegaSecondDerivative2} into
\eqr{EQomegaSecondDerivative1}, and grouping terms together allows to
formulate the following locally resolved two-body Noether sum rule:
\begin{align}
  \beta\langle \hat\Fv(\rv)\hat\Fv(\rv')  \rangle
  &=
  \Big\langle
  \frac{\delta^2 H[\eps]}{\delta\eps(\rv)\delta\eps(\rv')}
  \Big|_{\eps=0}  \Big\rangle.
  \label{EQsumRule2generic}
\end{align}
The correlator on the left hand side of \eqr{EQsumRule2generic} is
analogous to $\cov(\hat\Fv(\rv),\hat\Fv(\rv'))= \langle
\hat\Fv(\rv)\hat\Fv(\rv') \rangle$, because the local mean force
vanishes in equilibrium, $\langle\hat\Fv(\rv)\rangle=0$
\cite{hermann2022quantum, tschopp2022forceDFT}; we recall the
individual force contributions in the explicit
form~\eqref{EQforceDensityBalance} of the force balance relationship.

Equation \eqref{EQsumRule2generic} constitutes an exact sum rule that
relates the dyadic force-force correlations (left hand side) at two
different positions with the mean curvature of the Hamiltonian with
respect to local shifting (right hand side). Hence the left hand side
of this sum rule constitutes an immediately meaningful standalone
physical object, with clearcut statistical mechanical
interpretation. We demonstrate below that this is true also for the
more formal looking right hand side of \eqr{EQsumRule2generic}. That
the exact equality \eqref{EQsumRule2generic} holds generally, at
arbitrary positions $\rv$ and~$\rv'$, could have certainly not been
guessed easily a priori (we turn to partial integration methods below
and present details thereof in
Appendix~\ref{SECviaPartialIntegration}).

We re-write \eqr{EQsumRule2generic} via multiplying by~$\beta$ and
then treating the kinetic contributions separately (see
  Sec.~\ref{SECforceCorrelatorSplitting}). For brevity of notation it
is useful to combine the interparticle and external forces into the
following dedicated potential force density operator:
\begin{align}
  \hat\Fv_U(\rv) = \hat\Fv_{\rm
    int}(\rv)-\hat\rho(\rv)\nabla V_\ext(\rv),
  \label{EQFUoperator}
\end{align}
where the subscript $U$ refers to the total potential energy and we
recall the definition \eqref{EQFintOperator} of the interparticle
force operator $\hat\Fv_\rmint(\rv)$.  The total potential energy
operator is given as the phase space function $H_U = u(\rv^N)+\sum_i
V_\rmext(\rv_i)$.  Using the coordinate
transformation~\eqref{EQcanonicalTransformationCoordinates} the
potential force density is obtained via functional differentiation
according to $-\delta H_U/\delta\eps(\rv)|_{\eps=0}= \hat \Fv_U(\rv)$,
i.e.\ the first functional derivative,
cf.~\eqr{EQFoperatorFromDifferentiation}, of the potential energy
contribution to the Hamiltonian.

We identify distinct contributions (subscript~``dist'') that arise on
the left hand side of \eqr{EQsumRule2generic}. These terms are
generated from pairs of particles with unequal indices such that
double sums reduce to
$\sum_{ij(\neq)}\equiv\sum_{i=1}^N\sum_{j=1,j\neq i}^N$.  As we
  will show in detail below in Sec.~\ref{SECforceCorrelatorSplitting},
  we can then obtain from \eqr{EQsumRule2generic} the following pair
  of exact distinct and self identities:
\begin{align}
  &\avg{\beta\hat\Fv_U(\rv)\beta\hat\Fv_U(\rv')}_{\!\rm dist} =
  \nabla\nabla'\rho_2(\rv,\rv')
  \notag\\&\quad\qquad\qquad\quad
  + \avg{{\sum_{ij(\neq)}^{\phantom{space}}} 
    \delta(\rv-\rv_i)\delta(\rv'-\rv_j)
  \nabla_i\nabla_j \beta u(\rv^N)},
  \label{EQsumRuleFeFeDistinct}\\
  &\avg{\beta\hat\Fv_U(\rv)\beta\hat\Fv_U(\rv)}_{\!\rm self}
  =
  \nabla\nabla\rho(\rv) + \rho(\rv)\nabla\nabla\beta\Vext(\rv)
  \notag\\  &\qquad\quad 
  \qquad\quad \qquad\qquad+
  \avg{\sum_i \delta(\rv-\rv_i) \nabla_i\nabla_i\beta u(\rv^N)}.
  \label{EQsumRuleFeFeSelf}
\end{align}
The two-body density on the right hand side of
\eqr{EQsumRuleFeFeDistinct} consists of contributions only from
distinct particle pairs and it has the standard definition
\cite{hansen2013}:
\begin{align}
  \rho_2(\rv,\rv')
  &=\langle\hat\rho(\rv)\hat\rho(\rv')\rangle_{\rm
    dist}\notag\\
  &=
  \Big\langle\sum_{ij(\neq)}\delta(\rv-\rv_i)\delta(\rv'-\rv_j)\Big\rangle.
  \label{EQrho2distinctDefinition}
\end{align}
The self average on the left hand side of \eqr{EQsumRuleFeFeSelf} only
involves a single delta distribution in position, i.e., it is
defined such that
$\langle\beta\hat\Fv_U(\rv)\beta\hat\Fv_U(\rv)\rangle_{\rm self}
=\langle \sum_i [\nabla_i \beta u(\rv^N) + \nabla_i \beta
  V_\ext(\rv_i)][\nabla_i\beta u(\rv^N) + \nabla_i \beta
  V_\ext(\rv_i)] \delta(\rv-\rv_i) \rangle$.  
Thereby one can use
  $\delta(\rv-\rv_i)\delta(\rv'-\rv_i)=\delta(\rv-\rv')\delta(\rv-\rv_i)$
  inside of the integral over $\rv_i$ and then drop the common factor
  $\delta(\rv-\rv')$ on both sides of the self equation in order to
  arrive at the form \eqref{EQsumRuleFeFeSelf}.

The sum rules \eqref{EQsumRuleFeFeDistinct} and
\eqref{EQsumRuleFeFeSelf} give much concrete relevance to the more
abstract form \eqref{EQsumRule2generic}. In particular the curvature
terms on the right hand sides are given as explicit averages, which
are accessible in many-body simulations. The kinetic stress
autocorrelations that emerge from \eqr{EQsumRule2generic} are
addressed in Appendix~\ref{SECkineticEnergySingularity} and the energy
curvature contributions are discussed in
Appendix~\ref{SECkineticSelfCurvature}. We point the reader at precise
points in the following argumentation to these Appendices.

\section{Density-force Noether identity}
\label{SECdensityForceSumRule}

We next consider the correlator of the density operator with the total
force density operator, i.e.\ $\langle\hat\rho(\rv) \hat
\Fv(\rv')\rangle$.  Besides being interesting in its own right, this
correlator is relevant for the left hand side of
Eq.~\eqref{EQsumRule2generic} via the correlation of external and
total force densities, which can be written as
$-\langle\hat\rho(\rv)\hat\Fv(\rv') \rangle \beta \nabla
V_\rmext(\rv)$.

We demonstrate two distinct routes to formulate a valid sum rule for
the density-force correlator. First we start from the locally resolved
equilibrium force balance relationship \eqref{EQFisZero},
$\langle\hat\Fv(\rv)\rangle=0$. This identity holds regardless of the
form of the external potential $V_\rmext(\rv)$, and we can hence
functionally differentiate both sides of the equation with respect to
$V_{\rm ext}(\rv)$. Clearly the right hand side will remain
zero. Differentiating the left hand side yields
\begin{align}
  \frac{\delta\langle\hat\Fv(\rv)\rangle}
       {\delta V_{\rm ext}(\rv')} &=
  -\beta\cov(\hat\Fv(\rv),\hat\rho(\rv')) +
       \avg{\frac{\delta \hat\Fv(\rv)}{\delta V_{\rm ext}(\rv')}}.
       \label{EQdeltaForceDeltaVext}
\end{align}
That \eqr{EQdeltaForceDeltaVext} holds can be seen by writing out
explicitly the thermal average on the left hand side and functionally
differentiating all dependencies on the external potential as we will
show. As an aside, Eckert {\it et al.}~\cite{eckert2023fluctuation}
have recently pointed out that for an arbitrary phase space
function~$\hat A$, which is taken to be independent of
$V_\rmext(\rv)$, one has $\delta \langle \hat A\rangle/\delta
V_\rmext(\rv)=-\beta\cov(\hat A, \hat\rho(\rv))$, which upon choosing
$\hat A=\hat\Fv(\rv)$ generates the first term on the right hand side
of \eqr{EQdeltaForceDeltaVext}; the second term then stems from taking
account of the explicit dependence of $\hat \Fv(\rv)$ on
$V_\rmext(\rv)$ according to \eqr{EQForceDensityOperator} as we
demonstrate in the following. Considering the functional derivative on
the left hand side of \eqr{EQdeltaForceDeltaVext}, from the product
rule one obtains the following structure:
\begin{align}
  &  \Tr   \frac{\delta}{\delta V_\rmext(\rv')}  \Xi^{-1}
  e^{-\beta(H-\mu N)} \hat\Fv(\rv) \notag\\
  & \qquad = \Tr \frac{\delta \Xi^{-1}}{\delta V_\rmext(\rv')}
  e^{-\beta(H-\mu N)} \hat\Fv(\rv)
  \notag\\
  & \qquad \quad + \Tr \Xi^{-1} e^{-\beta(H-\mu N)}
  \Big[ \frac{\delta (-\beta H)}{\delta V_\rmext(\rv')} \hat\Fv(\rv)
    +\frac{\delta \hat\Fv(\rv)}{\delta V_\rmext(\rv')}
    \Big].
  \label{EQdeltaForceDeltaVextIntermediate}
\end{align}
The first term on the right hand side gives $\beta \rho(\rv')
\Fv(\rv)$ upon observing that $-\Xi^{-2} \delta \Xi/\delta
V_\rmext(\rv') = \Xi^{-1}\beta \rho(\rv')$, where the density profile
can be taken out of the trace in \eqr{EQdeltaForceDeltaVext}. The
second term follows straightforwardly as $-\beta\langle \hat\rho(\rv')
\Fv(\rv) \rangle + \langle \delta \hat \Fv(\rv)/\delta
V_\rmext(\rv')\rangle$ when taking account of the fact that $\delta
H/\delta V_{\rm ext}(\rv') = \hat\rho(\rv')$. Regrouping the terms
then yields \eqr{EQdeltaForceDeltaVext}.

The covariance on the right hand side of \eqr{EQdeltaForceDeltaVext}
reduces to the correlation, i.e., the mean of the product of the two
operators, $\cov(\beta\hat\Fv(\rv),\hat\rho(\rv'))=
\langle\beta\hat\Fv(\rv)\hat\rho(\rv')\rangle$, again because the
average force density vanishes in equilibrium,
$\langle\hat\Fv(\rv)\rangle=\Fv(\rv)=0$, cf.~\eqr{EQFisZero}. The
second term in Eq.~\eqref{EQdeltaForceDeltaVext} can be re-written as
follows:
\begin{align}
       \avg{\frac{\delta \hat\Fv(\rv)}{\delta V_{\rm ext}(\rv')}}
       &= -\avg{\frac{\delta}{\delta V_{\rm ext}(\rv')}
         \sum_i \delta(\rv-\rv_i) \nabla_i V_{\rm ext}(\rv_i)}\\
       &= -\avg{
         \sum_i\delta(\rv-\rv_i)\nabla_i \frac{\delta V_{\rm ext}(\rv_i)}
         {\delta V_{\rm ext}(\rv')}
       }\\
       &= \avg{\sum_i \delta(\rv-\rv_i) \nabla'\delta(\rv_i-\rv')
       }\\
       &= \nabla' \avg{\sum_i \delta(\rv-\rv_i)\delta(\rv_i-\rv')}\\
       &\equiv \nabla' \rho_2^\rmself(\rv,\rv').
\end{align}
In the last step above we have introduced the self two-body density,
defined as
\begin{align}
  \rho_2^{\rm self}(\rv,\rv') &= 
  \avg{\sum_i\delta(\rv-\rv_i)\delta(\rv'-\rv_i)}.
  \label{EQrho2selfDefinition}
\end{align}
In typical applications one would re-write the right hand side of
\eqr{EQrho2selfDefinition} as $\rho(\rv)\delta(\rv-\rv')$, but due to
the present gradient structure some care is required and we thus keep
$\rho_2^\rmself(\rv,\rv')$ in its full form
\eqref{EQrho2selfDefinition}.

Collecting terms and recalling that the right hand side of
Eq.~\eqref{EQdeltaForceDeltaVext} vanishes identically we obtain the
following compact force-density correlation sum rule:
\begin{align}
  \avg{\beta \hat \Fv(\rv) \hat\rho(\rv')} &=
  \nabla' \rho_2^\rmself(\rv,\rv').
  \label{EQdensityForceCorrelationSumRule}
\end{align}
Equation \eqref{EQdensityForceCorrelationSumRule} is a remarkable and
highly nontrival result, given the seemingly complex and {\it a
  priori} possibly highly correlated nature of its left hand
side. However, in reality only the relatively simple local term on the
right hand side remains. Hence for two distinct positions $\rv$ and
$\rv'$ the singular right hand side vanishes and
Eq.~\eqref{EQdensityForceCorrelationSumRule} reduces trivially to
\begin{align}
  \avg{\beta \hat \Fv(\rv) \hat\rho(\rv')} &= 0, \quad \rv\neq\rv'.
  \label{EQdensityForceCorrelationSumRuleDistinct}
\end{align}
We can also conclude, as the distinct contribution vansishes, that at
all positions $\rv$ and $\rv'$, even when $\rv=\rv'$, we have
\begin{align}
  \avg{\beta \hat \Fv(\rv) \hat\rho(\rv')}_{\rm dist} &= 0.
  \label{EQdensityForceCorrelationSumRuleDistinctAlterntive}
\end{align}

We recall that the present proof of the density-force sum rule
\eqref{EQdensityForceCorrelationSumRule} is based on functionally
differentiating $\Fv(\rv)=0$ with respect to $V_\rmext(\rv')$.  An
alternative derivation of Eq.~\eqref{EQdensityForceCorrelationSumRule}
can be based on the Noether invariance of the density profile. This
thermal symmetry is expressed as $\delta \rho(\rv')/\delta
\eps(\rv)|_{\eps=0}=0$. As before we denote the transformed positions
by $\tilde\rv_i$ and consider
\begin{align}
  \frac{\delta \rho(\rv')}{\delta\eps(\rv)} &=
  \frac{\delta}{\delta \eps(\rv)}
  \avg{\sum_i \delta(\rv'-\tilde\rv_i)}\\
  &=\avg{\beta\hat\Fv(\rv)\hat\rho(\rv')}
  +\avg{\sum_i\frac{\delta}{\delta \eps(\rv)}
    \delta(\rv'-\rv_i-\eps(\rv_i))},
  \label{EQdensityForceCorrelationIntermediate}
\end{align}
where the first term on the right hand side results from functionally
differentiating the probability distribution and it is already
evaluated at $\eps(\rv)=0$.  The second term can be re-written as
\begin{align}
  & \avg{
    \sum_i \nabla'
    \delta(\rv'-\rv_i-\eps(\rv_i))\cdot
    \frac{\delta [-\eps(\rv_i)]}{\delta \eps(\rv)}
  \Big|_{\eps=0}}
  \notag\\ &\qquad\qquad=
  -\nabla'\avg{\sum_i\delta(\rv'-\rv_i)\delta(\rv_i-\rv)}
  \notag\\ &\qquad\qquad=
  -\nabla'\rho_2^\rmself(\rv,\rv'),
  \label{EQgradRho2Result}
\end{align}
where we have used the definition \eqref{EQrho2selfDefinition} of the
self two-body density distribution $\rho_2^\rmself(\rv,\rv')$.  The
result \eqref{EQgradRho2Result}, when inserted into
\eqr{EQdensityForceCorrelationIntermediate} and recalling that the
left hand side thereof needs to vanish identically, gives the sum rule
\eqref{EQdensityForceCorrelationSumRule}.

That both derivations give identical results ultimately lies in the
fact that the order of differentation is the only genuine difference.
Starting from the free energy as an overarching object and building a
mixed second derivative, where the order of differentation is
interchanged yields: $\delta^2 \Omega[\eps]/[\delta \eps(\rv)\delta
  V_\ext(\rv')] =\delta^2 \Omega[\eps]/[\delta
  V_\ext(\rv')\delta\eps(\rv)]$ where we recall that
$\delta\Omega/\delta V_{\rm ext}(\rv)=\rho(\rv)$ and
$\delta\Omega[\eps]/\delta\eps(\rv)|_{\eps=0}= - \Fv(\rv)$ according
to \eqr{EQOmegaDerivativeLinear}.

Lastly, \eqr{EQdensityForceCorrelationSumRule} can be viewed as a
special case of the recent more general hyperforce correlation theory
by Robitschko {\it et al.}~\cite{robitschko2024any}. Their theory
applies to general phase space functions $\hat A(\rv^N,\pv^N)$ and for
configuration-dependent cases, $\hat A(\rv^N)$, it ascertains the
validity of the identity $\langle \beta \hat \Fv(\rv) \hat A(\rv^N)
\rangle=-\langle \sum_i\delta(\rv-\rv_i)\nabla_i \hat
A(\rv^N)\rangle$. Choosing the density operator as the observable of
interest, $\hat A(\rv^N)=\hat\rho(\rv')=\sum_j\delta(\rv'-\rv_j)$,
then replacing $\nabla_i\delta(\rv'-\rv_i)=
-\nabla'\delta(\rv'-\rv_i)$, and identifying
$\rho_2^\rmself(\rv,\rv')$ according to \eqr{EQrho2selfDefinition}
also yields \eqr{EQdensityForceCorrelationSumRule}.

In summary, we have obtained the density-force sum rule
\eqref{EQdensityForceCorrelationSumRule}, which is ready to use in the
further investigation of the local second-order Noether invariance
structure.  For subsequent use we find it convenient to re-write the
density-force correlation sum rule
\eqref{EQdensityForceCorrelationSumRule} using the splitting
\eqref{EQForceDensityOperator} of the total force density operator
$\hat\Fv(\rv)=\nabla\cdot\hat\taub(\rv)+\hat \Fv_U(\rv)$, where we
recall the definition~\eqref{EQkineticStressOperator} of the kinetic
stress density operator $\hat\taub(\rv)$ and the definition
\eqref{EQFUoperator} of the potential force density operator~$\hat
\Fv_U(\rv)$. After re-arranging \eqr{EQdensityForceCorrelationSumRule}
we find
\begin{align}
  \avg{\beta \hat\Fv_U(\rv)\hat\rho(\rv')}
  &= \nabla'\rho_2^\rmself(\rv,\rv')
  -\nabla\cdot\avg{\beta\hat\taub(\rv)\hat\rho(\rv')}
  \label{EQFUrhoFirst}\\
  &= (\nabla'+\nabla)\rho_2^\rmself(\rv,\rv')
  +\nabla \rho_2(\rv,\rv'),
  \label{EQFUrho}
\end{align}
where the form \eqref{EQFUrho} arises from performing the momentum
integrals over the standard Maxwell probability distribution in the
second term of \eqr{EQFUrhoFirst} and identifying $\langle
\hat\rho(\rv)\hat\rho(\rv')\rangle = \rho_2(\rv,\rv')
+\rho_2^\rmself(\rv,\rv')$.  Building furthermore the gradient with
respect to the position $\rv'$ and transposing [note that the
  transpose $(\nabla'\nabla)^{\sf T}=\nabla\nabla'$] yields
straightforwardly the following tensorial identities:
\begin{align}
   \avg{\beta\hat\Fv_U(\rv)\nabla'\hat\rho(\rv')} &=
  (\nabla'\nabla'+\nabla\nabla')\rho_2^\rmself(\rv,\rv')
  \notag\\&\quad
  +\nabla\nabla'\rho_2(\rv,\rv'),
  \label{EQFUgradrho}\\
  \avg{\nabla\hat\rho(\rv) \beta \hat \Fv_U(\rv')}
  &= (\nabla\nabla + \nabla\nabla')\rho_2^\rmself(\rv,\rv')
  \notag\\&\quad
  +\nabla\nabla'\rho_2(\rv,\rv').
  \label{EQgradrhoFU}
\end{align}
Equation \eqref{EQgradrhoFU} is thereby obtained from re-transposing
\eqref{EQFUgradrho} and interchanging the positions arguments $\rv$
and $\rv'$ upon observing the exchange symmetry
$\rho_2^\rmself(\rv,\rv')=\rho_2^\rmself(\rv',\rv)$; we recall the
definition \eqref{EQrho2selfDefinition} of the self two-body density
and clearly also $\rho_2(\rv,\rv')=\rho_2(\rv',\rv)$ as is apparent
from its definition \eqref{EQrho2distinctDefinition}.

The identities \eqref{EQFUgradrho} and \eqref{EQgradrhoFU} are now in
a form ready to be used subsequently in
Sec.~\ref{SECforceCorrelatorSplitting} in the proof of the Noether
two-body fore correlation sum rules \eqref{EQsumRuleFeFeDistinct} and
\eqref{EQsumRuleFeFeSelf}.

\section{Force correlator splitting}
\label{SECforceCorrelatorSplitting}

We aim to re-write the full force autocorrelator, as it e.g.\ appears
on the left hand side of \eqr{EQsumRule2generic}, via the
autocorrelator of the potential force density operator
$\hat\Fv_U(\rv)$, see \eqr{EQFUoperator}, in order to simplify the
structure of the trivial kinetic contributions. As above, we use the
definition \eqref{EQkineticStressOperator} of the kinetic stress
operator to express $\hat\Fv(\rv)=\nabla\cdot
\hat\taub(\rv)+\hat\Fv_U(\rv)$. We start with the force autocorrelator
and obtain by multiplying out and re-ordering:
\begin{align}
  \avg{\hat\Fv(\rv)\hat\Fv(\rv')} &=
  \avg{\nabla\cdot\hat\taub(\rv)\nabla'\cdot\hat\taub(\rv')}
  +  \avg{\hat\Fv_U(\rv)\hat\Fv_U'(\rv)}
  \notag\\
  &\;  +
  \avg{\hat\Fv_U(\rv)\nabla'\cdot\hat\taub(\rv')}
  + \avg{\nabla\cdot\hat\taub(\rv)\hat\Fv_U(\rv')}.
  \label{EQFeFePreliminary1}
\end{align}

We start with the purely kinetic two-body contribution, i.e.\ the
first term in \eqr{EQFeFePreliminary1} and obtain the following
result:
\begin{align}
 & \avg{\nabla\cdot\beta\hat\taub(\rv)
    \nabla'\cdot\beta\hat\taub(\rv')}\notag\\
  &= (\nabla\nabla'+\nabla'\nabla+\unity \nabla\cdot\nabla') 
  \rho_2^\rmself(\rv,\rv')
    +\nabla\nabla'\rho_2(\rv,\rv')
\label{EQdivTaudivTau}
\end{align}
where we have again used the definition \eqref{EQrho2selfDefinition}
of the self two-body density and we recall that~$\unity$ indicates the
$3\times 3$-unit matrix. In Appendix~\ref{SECkineticEnergySingularity}
we lay out the explicit calculation of both the self and distinct
contributions to \eqr{EQdivTaudivTau}.

The structure of the two mixed terms in \eqr{EQFeFePreliminary1}
allows to carry out the momentum integrals in a simple way. This
enables us to simplify the two kinetic contributions according to
\begin{align}
  \avg{\hat\Fv_U(\rv)\nabla'\cdot \beta\hat\taub(\rv')} &=
  -\avg{\hat\Fv_U(\rv)\nabla'\hat\rho(\rv')},\\
  \avg{\nabla\cdot\beta\hat\taub(\rv)\hat\Fv_U(\rv')} &=
  -\avg{\nabla\hat\rho(\rv) \hat\Fv_U(\rv')}.
\end{align}
  Thereby the momentum averages on the left hand side need to be
  carried out explicitly, upon using the Maxwell distribution, in
  order to simplify the divergence of the kinetic stress operator
  (left hand sides) to yield the gradient of the density operator
  (right hand sides). Then re-introducing the momentum average on the
  right hand side, as is implied by the $\langle\cdot\rangle$
  notation, creates no harm due to the independence of the operators
  on the right hand sides from the momentum degrees of freedom and the
  correct normalization of the average.

Collecting all terms and using Eqs.~\eqref{EQFUgradrho} and
\eqref{EQgradrhoFU} allows then to re-write \eqr{EQFeFePreliminary1}
upon multiplying by $\beta^2$ as
\begin{align}
  & \avg{\beta\hat\Fv(\rv)\beta\hat\Fv(\rv')} =
  \avg{\beta\hat\Fv_U(\rv)\beta\hat\Fv_U(\rv')} 
  -\nabla\nabla'\rho_2(\rv,\rv')
  \notag\\&\quad
  +(\unity \nabla\cdot\nabla' + \nabla'\nabla - \nabla\nabla'
  -\nabla'\nabla' - \nabla\nabla) \rho_2^\rmself(\rv,\rv').
  \label{EQFuFuvsFF}
\end{align}

Equation \eqref{EQFuFuvsFF} constitutes an explicit form of the left
hand side of the locally resolved two-body sum rule
\eqr{EQsumRule2generic}. We recall that the sum rule
\eqref{EQsumRule2generic} balances the force-force correlations on its
left hand side with an energy curvature term on the right hand side.

In order to address this right hand side, we split the Hamiltonian,
expressed in the new coordinates and hence functionally depending on
the shifting field $\eps(\rv)$, into kinetic and potential energy
contributions, $H[\eps]=H_{\rm kin}[\eps]+H_U[\eps]$.  The energy
curvature also consists of kinetic and potential energy contributions.
We defer the calculations for the kinetic energy curvature to
Appendix~\ref{SECkineticSelfCurvature} and for the potential energy
curvature to Appendix~\ref{SECpotentialEnergyCurvature}. The results
are as follows:
\begin{align}
  \avg{\frac{\delta^2H_{\rm kin}[\eps]}{\delta\eps(\rv)\delta\eps(\rv')}
  \Big|_{\eps=0}}
  &= k_BT (\unity \nabla\cdot\nabla' + 2\nabla'\nabla)
  \rho_2^\rmself(\rv,\rv'),
  \label{EQkineticEnergyCurvature}\\
  \avg{\frac{\delta^2 H_U[\eps]}{\delta\eps(\rv)\delta\eps(\rv')}\Big|_{\eps=0}}
  &=\notag\\&
  \avg{\sum_{ij}\delta(\rv-\rv_i)\delta(\rv'-\rv_j)\nabla_i\nabla_j u(\rv^N)}
  \notag\\&
  +\rho_2^\rmself(\rv,\rv')\nabla\nabla V_\rmext(\rv),
  \label{EQhurra}
\end{align}
and we recall that their sum constitutes the right hand side of the
generically expressed Nother sum rule \eqref{EQsumRule2generic}.

We can hence express \eqr{EQsumRule2generic} by rewriting its left
hand side via the right hand side of \eqr{EQFuFuvsFF} and its right
hand side via the sum of the right hand sides of the curvature results
\eqref{EQkineticEnergyCurvature} and \eqref{EQhurra}. Collecting all
terms yields the following result:
\begin{align}
  &  \avg{\beta\hat\Fv_U(\rv)\beta\hat\Fv_U(\rv')} - \nabla\nabla'\rho_2(\rv,\rv')
  \notag\\&\quad
  +(\unity\nabla\cdot\nabla'+\nabla'\nabla-\nabla\nabla'
  -\nabla'\nabla'-\nabla\nabla)\rho_2^\rmself(\rv,\rv')=
  \notag\\&
  (\unity\nabla\cdot\nabla'+2\nabla'\nabla)\rho_2^\rmself(\rv,\rv')
  \notag\\&\quad
  +\avg{\sum_{ij}\delta(\rv-\rv_i)\delta(\rv'-\rv_j)\nabla_i\nabla_j \beta u(\rv^N)}
  \notag\\&\quad
  +\rho_2^\rmself(\rv,\rv')\nabla\nabla \beta V_\rmext(\rv).
\end{align}
Re-arranging and simplifying yields:
\begin{align}
  &\avg{\beta\hat\Fv_U(\rv)\beta\hat\Fv_U(\rv')} = 
  \notag\\&\quad
  \nabla\nabla'\rho_2(\rv,\rv')
  +\delta(\rv-\rv')[\nabla\nabla\rho(\rv) 
+\rho(\rv)\nabla\nabla \beta V_\rmext(\rv)]
  \notag\\&\quad
  +\avg{\sum_{ij}\delta(\rv-\rv_i)\delta(\rv'-\rv_j)
    \nabla_i\nabla_j\beta u(\rv^N)},
  \label{EQFUFUfromExplicitCalculation}
\end{align}
where we could finally replace the self two-body density as
follows. In the external curvature term we have
$\rho_2^\rmself(\rv,\rv')=\delta(\rv-\rv')\rho(\rv)$ and in the ideal
contribution we could simplify $(\nabla+\nabla')(\nabla+\nabla')
\rho_2^\rmself(\rv,\rv')= \delta(\rv-\rv')\nabla\nabla\rho(\rv)$,
which can be verified explicitly by calculating derivatives in
$(\nabla+\nabla')(\nabla+\nabla')[\delta(\rv-\rv')\rho(\rv)]$ using
the product rule and exploiting
$\nabla'\delta(\rv-\rv')=-\nabla\delta(\rv-\rv')$.

  One can split \eqr{EQFUFUfromExplicitCalculation} into regular
  and singular contributions, where the latter are identified as
  having a common factor $\delta(\rv-\rv')$. This splitting
  discriminates between self and distinct contributions on the basis
  of the standard criterion for pairs of particle indices which are
  equal $i=j$ (self) and different $i\neq j$ (distinct). Hence
  splitting \eqr{EQFUFUfromExplicitCalculation} leads respectively to
  the distinct sum rule \eqref{EQsumRuleFeFeDistinct} and, upon
  leaving away the common factor $\delta(\rv-\rv')$ from the self
  terms, to the self sum rule \eqref{EQsumRuleFeFeSelf}.

This completes our proof of the locally resolved two-body Noether
invariance sum rules \eqref{EQsumRuleFeFeDistinct} and
\eqref{EQsumRuleFeFeSelf}. As mentioned above, all derivations
continue to hold in the canonical ensemble. The mechanism for this
universality is the primarily mechanical nature, we recall the
shifting transformation \eqref{EQcanonicalTransformationCoordinates}
and \eqref{EQcanonicalTransformationMomenta}, of the considered
thermal invariance, which is oblivious to the presence of a particle
bath.

Our theory is now ready to be applied to real systems and we present
explicit results for a variety of common soft matter models in
Sec.~\ref{SECbulkLiquidNoetherStructure}. These results demonstrate
the validity of the sum rules and they show the prowess of the
force-force and force-gradient correlation functions to systematically
quantify and shed light on the self-structuring mechanisms at play in
equilibrium.  Readers who are keen to first follow further formal
argumentation that demonstrates the validity of the Nother sum rules
are welcome to go to Appendix~\ref{SECviaPartialIntegration}, where we
lay out our partial phase space integration route.

\begin{figure*}[htb!]
    \includegraphics[width=0.79\linewidth]{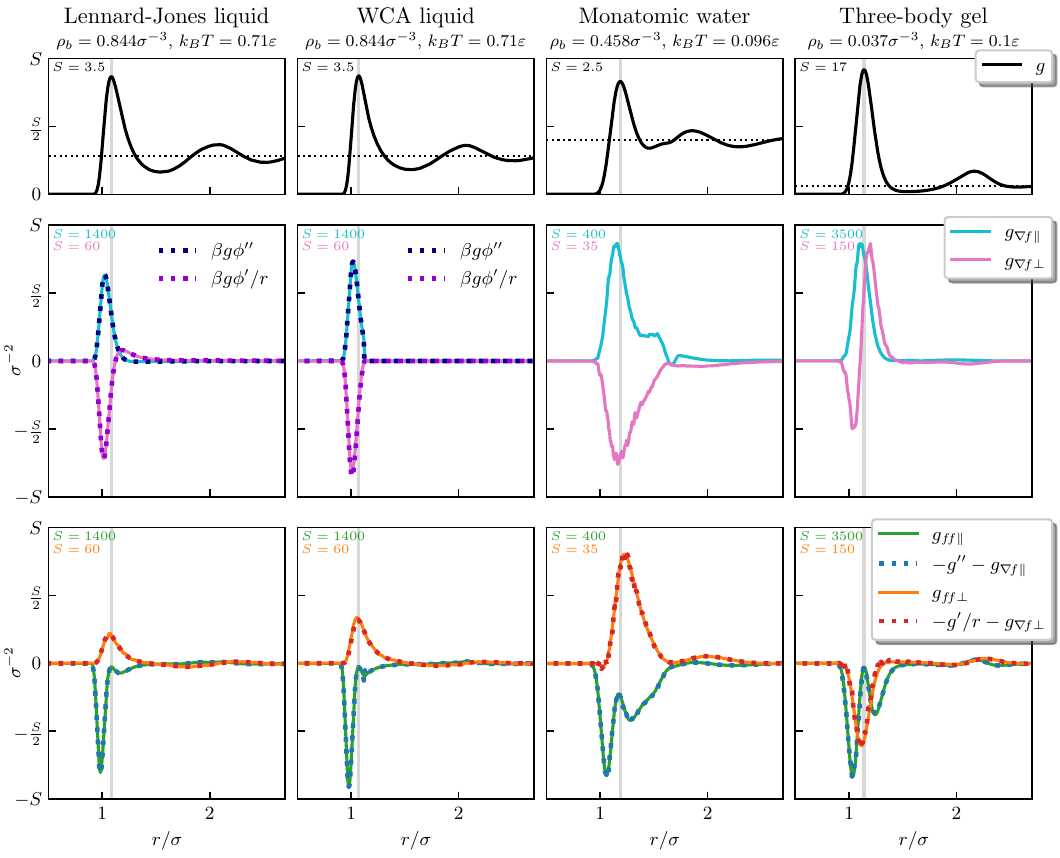}
\caption{Correlation functions analogous to Fig.~\ref{FIG1}, with
  results for the Lennard-Jones liquid shown again as a reference
  (first column), the Weeks-Chandler-Andersen fluid (second column),
  monatomic water at (approximately) ambient conditions
  \cite{molinero2009, coe2022water} (third column), and the three-body
  interacting gel \cite{saw2009, saw2011, sammueller2023gel} (fourth
  column); our model parameter choices are inline with the values
    given in these references. Shown are results for the standard
  pair correlation function $g(r)$ for each system (top row).  Further
  two-body structure is revealed by the radial ($\parallel$) and
  transversal ($\perp$) components of the force-gradient correlator
  ${\sf g}_\gradf(r)$ (middle row) and of the force-force correlator
  ${\sf g}_\ff(r)$ (bottom row). The vertical gray lines indicate
    the position of the first maximum of $g(r)$ as a guide to the
    eye. The dotted lines in the middle row indicate the two
  components of the force-gradient correlation function as obtained
  from the simplifications \eqref{EQsimpleFluidParallel} and
  \eqref{EQsimpleFluidPerpendicular} for the two pairwise
  (Lennard-Jones and Weeks-Chandler-Andersen) Hamiltonians. The
    agreement with the respective solid lines, as obtained from
    numerically differentiating the interparticle forces, demonstrates
    that the force gradient correlation function follows directly from
    the product of $g(r)$ and the scaled derivatives $\beta\phi''(r)$
    and $\beta\phi'(r)/r$ of the pair potential; see
    Eqs.~\eqref{EQsimpleFluidParallel} and
    \eqref{EQsimpleFluidPerpendicular}.  The dotted lines in the
  bottom row indicate the results according to the Noether force sum
  rules \eqref{EQgIdentityManyBodyParallel} and
  \eqref{EQgIdentityManyBodyPerpendicular}. The agreement with
    the respective solid lines demonstrates the validity of these sum
    rules. Taken from Ref.~\cite{sammueller2023whatIsLiquid}.
  \label{FIG2}
  }
\end{figure*}
\begin{figure*}[htb!]
    \includegraphics[width=0.99\linewidth]{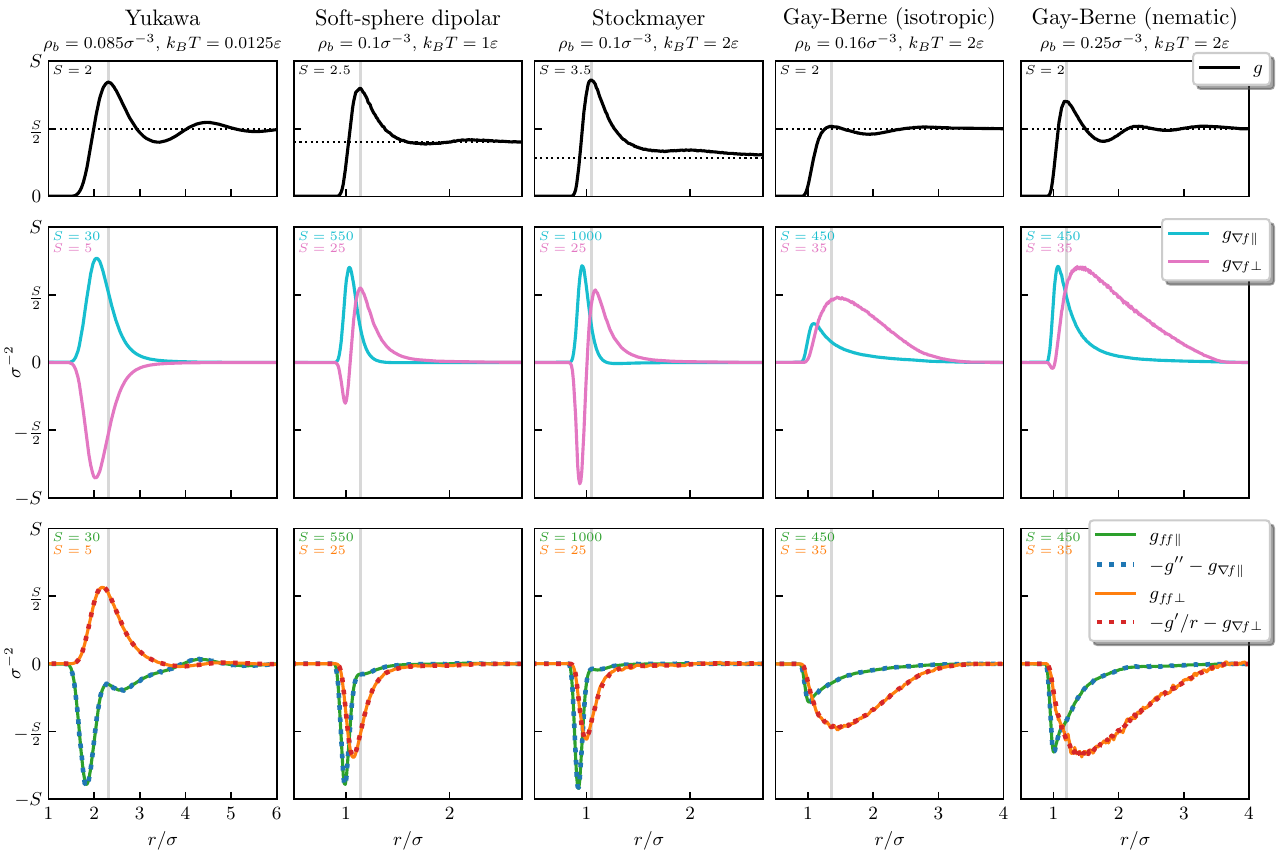}
\caption{Correlation functions analogous to those shown in
  Figs.~\ref{FIG1} and \ref{FIG2}, but for the Yukawa liquid (first
  column), the soft-sphere dipolar fluid (second column), the
  Stockmayer fluid (third column), and the Gay-Berne model in the
  isotropic (fourth column) and nematic phase (fifth column). The
  results for the anisotropic models are obtained from canonical Monte
  Carlo simulations, and they are averaged over the microscopic
  orientations; the simulation box volume is $V=(20\sigma)^3$ and the
  long-ranged interactions are cut off at radial distance of
  $10\sigma$.  Shown are the pair correlation function $g(r)$ (top
  row), the force-gradient correlator ${\sf g}_\gradf(r)$ (middle
  row), and the force-force correlator ${\sf g}_\ff(r)$ together with
  the result from the Noether identities
  \eqref{EQgIdentityManyBodyParallel} and
  \eqref{EQgIdentityManyBodyPerpendicular} (bottom row). The
    excellent agreement of the pairs of dotted and solid lines in all
    panels in the bottom row demonstrates the universal validity of
    these sum rules.
    Parameter choices are inverse screening parameter
    $\kappa=2/\sigma$ for the Yukawa liquid, dipolar strength
    $\mu/\sqrt{\epsilon\sigma^3}=2$ for both the soft-sphere dipolar
    fluid and the Stockmayer fluid, and $\kappa = 3.8$ and $\kappa' =
    5$ for the Gay-Berne model. Adapted from
  Ref.~\cite{sammueller2023whatIsLiquid} (Supplementary Material).
   \label{FIG3}
  }
\end{figure*}

\section{Noether structure in bulk}
\label{SECbulkLiquidNoetherStructure}

We specialize to homogeneous bulk liquid states, where
$\rho(\rv)=\rho_b=\rm const$ and $V_\ext(\rv)=0$. The potential forces
then only arise from interparticle contributions and hence
$\hat\Fv_U(\rv)=\hat\Fv_{\rm int}(\rv)$ with the interparticle force
density operator $\hat\Fv_\rmint(\rv)$ being given by
\eqr{EQFintOperator}. We address the distinct sum rule
\eqref{EQsumRuleFeFeDistinct} and use the standard pair correlation
function or ``radial distribution function'' $g(r)$. Here
$r=|\rv-\rv'|$ denotes the separation distance between the two
particles.  The relationship to the two-body density is simply via
normalization with the squared bulk density~$\rho_b$:
\begin{align}
  g(|\rv-\rv'|)=\rho_2(\rv,\rv') /\rho_b^2.
  \label{EQgofrDefinition}
\end{align}
Furthermore we follow Ref.~\cite{sammueller2023whatIsLiquid} in
introducing both the force-force pair correlation function ${\sf
  g}_\ff(r)$ and the force gradient correlator ${\sf
  g}_\gradf(r)$. These are rank-two tensorial correlation functions,
as represented by $3\times 3$-matrices in case of a three-dimensional
system, and they are defined via
\begin{align}
  {\sf g}_\ff(|\rv-\rv'|) &= \frac{\beta^2}{\rho_b^2} \langle\hat\Fv_{\rm
    int}(\rv)\hat\Fv_{\rm int}(\rv')\rangle_{\rm dist},
  \label{EQgffDefinition}\\
  {\sf g}_\gradf(|\rv-\rv'|)&=
  \notag\\
  -\frac{\beta}{\rho_b^2}\Big\langle
  \sum_{ij(\neq)}
  &
  \delta(\rv-\rv_i)\delta(\rv'-\rv_j)\nabla_i\nabla_j
  u(\rv^N) \Big\rangle.
  \label{EQggradfDefinition}
\end{align}
The force gradient correlator ${\sf g}_\gradf(r)$ is also the negative
mean potential curvature, as is apparent from the right hand side of
\eqr{EQggradfDefinition} via $\nabla_i\nabla_j u(\rv^N)$.  One can
hence refer to the correlator also as the mean negative Hessian, with
respect to differentiation by particle positions, of the interparticle
potential energy. 

On the other hand an equally valid interpretation is that of the
negative interparticle force gradient. Recalling the interparticle
force on particle $i$ as $\fv_i(\rv^N)=-\nabla_i u(\rv^N)$ and that on
particle $j$ as $\fv_j(\rv^N)=-\nabla_j u(\rv^N)$ we can re-write the
potential energy curvature as $\nabla_i\nabla_j u(\rv^N) = -\nabla_i
\fv_j(\rv^N) = -[\nabla_j \fv_i(\rv^N)]^{\sf T}$, where the
possibility of interchange of the particle indices in the two latter
expressions is due to interchange of the order of the two derivatives.

Expressing the forces in this way allows to re-write
Eqs.~\eqref{EQgffDefinition} and \eqref{EQggradfDefinition} in the
following more explicit forms:
\begin{align}
  &{\sf g}_\ff(|\rv-\rv'|)=
  \notag\\&\qquad 
  \avg{\sum_{ij(\neq)}\delta(\rv-\rv_i)\delta(\rv'-\rv_j)
    \beta\fv_i(\rv^N)\beta\fv_j(\rv^N)}\Big/\rho_b^2,
  \label{EQforceForceTwoPoint}\\
  &{\sf g}_\gradf(|\rv-\rv'|)=
  \notag\\&\qquad
  \avg{\sum_{ij(\neq)}\delta(\rv-\rv_i)\delta(\rv'-\rv_j)\nabla_i 
    \beta\fv_j(\rv^N)}\Big/\rho_b^2,
  \label{EQforceGradientTwoPoint}
\end{align}
where in the last term one can alternatively replace $\nabla_i
\fv_j(\rv^N)=[\nabla_j \fv_i(\rv^N)]^{\sf T}$, as laid out above.

In the fully position-resolved case, which is appropriate for the
investigation of spatially inhomogeneous situations, where
$\rho(\rv)\neq \rm const$, suitable correlators are the following
two-body force-force and force-gradient density distributions:
\begin{align}
  &{\sf G}_\ff(\rv,\rv')=
  \avg{\sum_{ij(\neq)}\delta(\rv-\rv_i)\delta(\rv'-\rv_j)
    \beta\fv_i(\rv^N)\beta\fv_j(\rv^N)},
  \label{EQGforceForceTwoPoint}\\
  &{\sf G}_\gradf(\rv,\rv')=
  \avg{\sum_{ij(\neq)}\delta(\rv-\rv_i)\delta(\rv'-\rv_j)\nabla_i 
    \beta\fv_j(\rv^N)}.
  \label{EQGforceGradientTwoPoint}
\end{align}
Inhomogeneous versions of Eqs.~\eqref{EQforceForceTwoPoint} and
\eqref{EQforceGradientTwoPoint} can be obtained via normalization with
the (inhomogeneous) density profile: ${\sf
  G}_\ff(\rv,\rv')/[\rho(\rv)\rho(\rv')]$ and ${\sf
  G}_\gradf(\rv,\rv')/[\rho(\rv)\rho(\rv')]$.

As announced in the introduction, using the two-body correlators
defined via Eqs.~\eqref{EQgofrDefinition}-\eqref{EQggradfDefinition}
the distinct sum rule \eqref{EQsumRuleFeFeDistinct} can be written for
the case of a bulk fluid in the following appealing compact form
\cite{sammueller2023whatIsLiquid}:
\begin{align}
  \nabla\nabla g(r) & + {\sf g}_\gradf(r) + {\sf g}_\ff(r) = 0,
  \label{EQgIdentity}
\end{align}
which is Eq.~\eqref{EQ3gRuleIntroduction}.  Due to the rotational
symmetry of a bulk fluid, the only nontrivial tensor components are
parallel ($\parallel$) and transversal ($\perp$) to $\rv-\rv'$, such
that \eqr{EQgIdentity} reduces to the following two nontrivial tensor
components:
\begin{align}
  g''(r) + g_{\gradf\parallel}(r) + g_{\ff\parallel}(r)   &= 0,
  \label{EQgIdentityManyBodyParallel}\\
   \frac{g'(r)}{r} + g_{\gradf\perp}(r) + g_{\ff\perp}(r)  &= 0,
  \label{EQgIdentityManyBodyPerpendicular}
\end{align}
with the prime denoting the derivative with respect to~$r$. In the
chosen coordinate system the matrices are diagonal, ${\rm
  diag}(\parallel,\perp,\perp)$, with the first axis being parallel to
$\rv-\rv'$ and the second and third components being equal and
corresponding to two mutually perpendicular directions which are also
perpendicular to $\rv-\rv'$.

More explicitly, in the considered radial symmetry the second
derivative operator in \eqr{EQgIdentity} reduces to $\nabla\nabla =
\ev_\parallel\ev_\parallel \partial^2/\partial r^2 + (\unity -
\ev_\parallel\ev_\parallel) r^{-1}\partial/\partial r$, where the
direction that connects the two points is
$\ev_\parallel=(\rv-\rv')/|\rv-\rv'|$. This unit vector is
complemented by two orthogonal directions $\ev_\perp,\ev_\perp'$ such
that all three $\ev_\parallel,\ev_\perp,\ev_\perp'$ are mutually
orthogonal to each other. The perpendicular components then generate
the $g'(r)/r$ term in \eqr{EQgIdentityManyBodyPerpendicular}.  These
tensor operations can be efficiently performed in numerical work and
we give details about the sampling strategies used in our simulations
in Appendix \ref{SECappendixMeasuring}.

For simple fluids, with the particles interacting mutually only via a
pair potential $\phi(r)$, the force gradient correlator
\eqref{EQggradfDefinition} reduces to ${\sf g}_\gradf(r)=\beta
g(r)\nabla\nabla \phi(r)$ such that the two non-trivial components
($\parallel$ and $\perp$) can be written according to
\begin{align}
 g_{\gradf\parallel}(r) &= \beta g(r)\phi''(r),
  \label{EQsimpleFluidParallel}\\
 g_{\gradf\perp}(r) &= \beta g(r)\frac{\phi'(r)}{r}.
  \label{EQsimpleFluidPerpendicular}
\end{align}
This simplification is due to the reduction of the mixed derivative
$\nabla_i\nabla_j u(\rv^N)=\nabla_i\nabla_j
\sum_{kl(\neq)}\phi(|\rv_k-\rv_l|)/2= \nabla_i \nabla_j
\phi(|\rv_i-\rv_j|)$, for $i\neq j$, such that strikingly only a
single pair potential bond, that between the particle $i$ and $j$,
contributes and the sums over particles $k$ and $l$ have disappeared.
 The fact that the separation distance~$r$ between the two
  particles is fixed via the delta distributions in position allows to
  take the Hessian $\nabla\nabla\phi(r)$ outside of the average, with
  the remaining average then generating the bare pair correlation
  function $g(r)$; we recall its definition via
  Eqs.~\eqref{EQrho2distinctDefinition} and \eqref{EQgofrDefinition}.

As a consequence of the structure of the curvature correlator
expressed in Eqs.~\eqref{EQsimpleFluidParallel} and
\eqref{EQsimpleFluidPerpendicular}, we can rewrite the sum rules
\eqref{EQgIdentityManyBodyParallel} and
\eqref{EQgIdentityManyBodyPerpendicular} for the case of pair-wise
interparticle forces in the following form:
\begin{align}
  g''(r) + \beta\phi''(r) g(r) + g_{\ff\parallel}(r)=0,
  \label{EQsumRuleSimple1}\\
  \frac{g'(r)}{r} + \frac{\beta\phi'(r)}{r}g(r) + g_{\ff\perp}(r)=0.
  \label{EQsumRuleSimple2}
\end{align}
The validity of the identities \eqref{EQsumRuleSimple1} and
\eqref{EQsumRuleSimple2} can be analytically verified in the
low-density limit of a simple fluid, where $g(r)=\exp(-\beta\phi(r))$,
as laid out in Ref.~\cite{sammueller2023whatIsLiquid}. We here
reproduce the argument.  In the low-density limit the force-force
correlations are due to the antiparallel direct forces between a
particle pair:
$g_{\ff\parallel}(r)=-g(r)[\beta\phi'(r)]^2$. Furthermore
$g_{\ff\perp}(r)=0$ due to the absence of a third particle at
$\rho_b\to 0$ that could mediate a transversal force.  Insertion into
Eqs.~\eqref{EQsumRuleSimple1} and \eqref{EQsumRuleSimple2} verifies
this solution.

As an illustration of the validity of these sum rules and for
demonstrating that fresh insight into the force correlation structure
can be obtained \cite{sammueller2023whatIsLiquid}, we return to the
Lennard-Jones system mentioned in the Introduction. This fundamental
model for simple systems comprises gas, liquid and crystal
phases. While our considerations were specific to the symmetry of
fluid phases, the structure of Eqs.~\eqref{EQsumRuleSimple1} and
\eqref{EQsumRuleSimple2} remains intact upon averaging the distance
vector $\rv-\rv'$ over all orientations and translations.  We recall
the display of simulation results in Fig.~\ref{FIG1} and in
  particular the validity of the sum rules
  \eqref{EQgIdentityManyBodyParallel} and
  \eqref{EQgIdentityManyBodyPerpendicular}, indicated by the agreement
  of the solid and dashed lines in the bottom row of Fig.~\ref{FIG1}.
The parallel component of the force gradient correlator
$g_{\gradf\parallel}(r)$ has a strong positive peak at a distance near
the Lennard-Jones diameter. The perpendicular component
$g_{\gradf\perp}(r)$ is negative, which indicates anti-correlation as
is mediated by the surrounding particles. The structure of the crystal
is particularly rich, as can be expected. In all phases the sum rules
are satisfied to high numerical precision, which is remarkable as the
three correlation functions $g(r)$, $\gff(r)$, and $\ggradf(r)$ seem
to be {\it a priori} independent of each other.

We show results for the Weeks-Chandler-Andersen liquid, for monatomic
water, using the parameterization by Molinero and Moore
\cite{molinero2009} of the Stillinger-Weber potential
\cite{stillinger1985}, and for the three-body interacting gel
\cite{saw2009, saw2011, sammueller2023gel} in Fig.~\ref{FIG2};
  the model parameter values are inline with the choices in these
  references. The results for the Lennard-Jones liquid are shown as a
reference. Going from the Lennard-Jones liquid (first column) to the
Weeks-Chandler-Andersen liquid (second column) at identical
thermodynamic conditions is a setup that lets one assess the influence
of the interparticle attractive forces, which are absent in the latter
model. As is well-known, the pair correlation function $g(r)$ is only
very mildly affected by interparticle attraction. In striking contrast
the clear positive peak of the Lennard-Jones liquid in the transversal
mean force gradient $g_{\gradf\perp}(r)$ is absent in the
Weeks-Chandler-Andersen liquid.
 This behaviour is very notable due to the common view of
  interparticle attraction having no significant effect on the
  microscopic structure of liquids. While our present model comparison
  explicitly confirms this expectation based solely on the behaviour
  of the pair correlation function $g(r)$, a much more complete
  picture emerges from the full force-based pair correlation
  structure. As it turns out, $g_\gradf(r)$ is a suitable means to
  clearly point to the presence of interparticle attraction. We recall
  the relationship \eqref{EQsimpleFluidPerpendicular} with the pair
  potential.

Going to the monatomic Molinero-Moore \cite{molinero2009,
  coe2022water} water model (third column) reveals richer force
gradient and force-force correlation structure over a broader range of
distances $r$ as compared to the previous pairwise models. We recall
that the three-body gel \cite{saw2009, saw2011, sammueller2023gel} is,
similar to the monatomic water model, a mere reparametrization of the
original Stillinger-Weber \cite{stillinger1985} model. Yet both the
force-gradient and the force-force correlation structure (fourth
column) changes dramatically when going from the liquid to the
gel. The perpendicular mean force-gradient $g_{\gradf\perp}(r)$
develops a strong positive signal immediately beyond the first peak of
$g(r)$, which is indicated by the vertical gray line as a
  reference.

Even more strikingly the transverse force-force correlation function
$g_{\ff\perp}(r)$ has a prominent negative peak as opposed to the
positive peak observed in all considered liquids. Such negative
force-force correlations are indicative of mechanical stability of
particle strands that the gel former develops and which are very
apparent in simulation snapshots (see
e.g.\ Ref.~\cite{sammueller2023gel}). These features cannot be
discerned from $g(r)$ alone, which despite its quantitatively
exaggerated appearance remains liquid-like, cf.\ the top right panel
in Fig.~\ref{FIG2}.

In Fig.~\ref{FIG3} we show results for the Yukawa liquid, the
soft-sphere dipolar fluid, the Stockmayer fluid
\cite{allen2019topRev}, and the Gay-Berne model \cite{gay1981,
  brown1998, allen2019topRev} in the isotropic and nematic phase. Here
we use Monte Carlo simulations to generate the numerical results, as
this method is more straightforward to use for the present models with
orientational degrees of freedom. Choices for parameter values
  are given in the caption of Fig.~\ref{FIG3}, following
  Refs.~\cite{stevens1995, brown1998}. In all cases considered the
sum rules were found to be satisfied to high numerical accuracy.

The results for the point-Yukawa fluid
\cite{hamaguchi1997,dijkstra1998yukawa} (first column) serve again to
illustrate generic liquid-like structuring. No positive signal in
$g_{\gradf\perp}(r)$ occurs as is consistent with the purely repulsive
character of the model and the correlation structure is spread out
over a larger range of distance as opposed to the Lennard-Jones liquid
(recall the first column of Fig.~\ref{FIG2}), which is consistent with
the longer-ranged decay of the Yukawa pair potential. Both the
soft-sphere dipolar system (second column) and the Stockmayer model
(third column) display prominently the signs of chain formation that
we had previously identified for the colloidal gel former: The
perpendicular mean force gradient $g_{\gradf\perp}(r)$ has a strong
positive peak directly beyond the first peak of $g(r)$ and the
prominent peak of $g_{\ff\perp}(r)$ is negative rather than positive
as we find it to be in the simple fluid phases and in monatomic
(liquid) water.
 
Addressing the liquid-crystal forming Gay-Berne model reveals further
differences. Even the isotropic phase (fourth column of
Fig.~\ref{FIG3}) displays pronounced differences to all previous
models, in that $g_{\gradf\perp}(r)$ has lost its negative peak at
distances smaller than the location of the first peak of $g(r)$
  (vertical gray line).  Instead $g_{\gradf\perp}(r)$ displays a
prominent entirely positive peak, while the behaviour of both
force-force correlation components is qualitatively similar to the
chain forming models. Curiously, upon increasing the density, such
that the Gay-Berne model spontaneously develops nematic order,
pronounced quantitative increases occur in all observed features of
both the mean force-gradient and the force-force correlator. We
re-iterate that we have deliberately averaged over all orientations
for simplicity. We expect that resolving the dependence on orientation
with respect to the nematic director can reveal much further insight
into the spatial structure of liquid crystal ordering.

\section{Conclusions}
\label{SECconclusions}

In conclusion we have provided a detailed description of the recent
Noether invariance theory for the two-body structure of liquids and of
more general soft matter systems \cite{sammueller2023whatIsLiquid}.
The theory is based on a specific spatial displacement invariance of
the fundamental degrees of freedom of the statistical many-body
physics, cf.~Eqs.~\eqref{EQcanonicalTransformationCoordinates}
and~\eqref{EQcanonicalTransformationMomenta}. Originally formulated as
a global operation \cite{hermann2021noether, hermann2022topicalReview,
  hermann2022variance} and subsequently generalized to local shifting
\cite{tschopp2022forceDFT, hermann2022quantum}, the approach yields
new insight into the correlated pair structure of soft matter
\cite{sammueller2023whatIsLiquid}.

We have here provided a detailed account of the theory, giving
derivations, including the treatment of all self-contributions as well
as alternative routes of proof explicitly. We have carefully treated
the contributions that arise from the kinetic energy part of the
Hamiltonian and have shown how these reduce in equilibrium to simple
diffusive terms. As illustrations of the framework, we have discussed
computer simulation results for the Lennard-Jones system in all its
three phases, for the Yukawa liquid, a soft-sphere dipolar fluid, the
Stockmayer fluid, as well as the Gay-Berne model in the isotropic and
the nematic phase. We have also shown results for the
Weeks-Chandler-Andersen liquid, for monatomic water
\cite{molinero2009, coe2022water}, and for the many-body colloidal gel
former \cite{saw2009, saw2011, sammueller2023gel}.

The results indicate that the force-based pair correlation functions
reveal much additional insight into the spatial structure over what is
provided by $g(r)$ alone. Crucially, neither their computational cost
in simulations nor the graphical representations as mere
distance-dependent plots pose any practical difficulties.  In our
present investigation the computationally most expensive simulation
procedure is to calculate the force gradients of the three-body
interacting Hamiltonians via numerical finite differences. As we have
shown it can be sufficient to only consider the two relevant tensor
components, i.e., the radial and transversal directions relative to
the pair distance vector.  Hence in terms of simplicity, much of the
appeal of $g(r)$ is retained, while the basis in the Noether
invariance also makes for a well-grounded statistical mechanical
foundation.

Our results add to the body of sum rules in statistical mechanics
\cite{baus1984, evans1990, henderson1992, triezenberg1972} and they
share formal similarities with sum rules for interface Hamiltonians
\cite{mikheev1991}, see Ref.~\cite{squarcini2022} for recent work, and
with the Takahashi-Ward identities \cite{takahashi1957, ward1950} of
quantum field theory. As locally resolved interparticle force
measurements have been demonstrated in colloidal systems
\cite{dong2022}, future experimental use of the Noether force
correlation functions looks feasible. Relating to force-sampling
methods that reduce the statistical variance inherent in sampling
results \cite{borgis2013, rotenberg2020, delasheras2018forceSampling,
  purohit2019, coles2019, coles2021, coles2023revelsMD} is another
interesting point for future work, as can be relating to force-based
density functional theory
\cite{tschopp2022forceDFT,sammueller2022forceDFT}.  The recent neural
functional theory \cite{delasheras2023perspective,
  sammueller2023neural, sammueller2023whyNeural,
  sammueller2023neuralTutorial} is based on the powerful concept of
using neural networks to represent functional relationships which
encapsulate the correlation behaviour of complex systems. Noether sum
rules have been shown to provide valuable consistency checks for these
neural functionals and they give much inspiration for further
theoretical developments in the spirit of physics-informed machine
learning \cite{karniadakis2021, ciarella2022, janzen2023,
  jung2023piml, rodrigues2023, wu2023review}.

For the reason of being fully explicit, we have formulated the theory
in the grand ensemble, as is certainly common for the present type of
considerations. However, the entirety of the argumentation is
applicable to a canonical treatment, where the particle number is
fixed. A simple way to see the equivalence is to recognize that we
have never relied on or indeed exploited the grand ensemble structure;
no derivatives with respect to $\mu$ are taken or similar operations
being carried out. Hence the Noether correlation theory does not
interfere with approaches that specifically target on particle number
effects, such as the local compressibility introduced by Evans and
coworkers \cite{evans2015jpcm, evans2019pnas, coe2022prl}.  The
Noether approach captures genuinely the mechanical fluctuations, as
opposed to local chemical (particle number) \cite{evans2015jpcm,
  evans2019pnas, coe2022prl, eckert2020, eckert2023fluctuation} and
thermal (energy) fluctuations \cite{coe2022prl, eckert2020,
  eckert2023fluctuation}.  We lastly point to the recent hyperforce
generalization \cite{robitschko2024any} that arises from thermal
Noether invariance and that applies to arbitrary observables.

\acknowledgments We thank Silas Robitschko and Daniel de las Heras for
useful discussions.

\appendix

\section{Kinetic stress autocorrelation}
\label{SECkineticEnergySingularity}

The Noether identity \eqref{EQsumRule2generic} contains a singular
self contribution which arises from the autocorrelation of the kinetic
stress \eqref{EQkineticStressOperator} on the left hand side and from
the kinetic energy on the right hand side. We treat this singular
contribution to the sum rule in the following by first considering the
kinetic stress autocorrelator, as it is contained in the form of a
force autocorrelator on the left hand side of \eqr{EQsumRule2generic}
and correspondingly occuring (multiplied by $\beta$) on the left hand
side of \eqr{EQdivTaudivTau}, i.e.,
\begin{align}
  \beta^2\Big\langle
  \nabla\cdot\sum_i\frac{\pv_i\pv_i}{m}\delta(\rv-\rv_i)
  \nabla'\cdot\sum_j\frac{\pv_j\pv_j}{m}\delta(\rv'-\rv_j)
  \Big\rangle.
  \label{EQkineticSingular1}
\end{align}
The self part of \eqr{EQkineticSingular1} consists only of the cases
$i=j$ in the double sum. We can hence simplify
\eqr{EQkineticSingular1} according to:
\begin{align}
  \nabla\nabla':\beta^2\Big\langle
  \sum_i \frac{\pv_i\pv_i\pv_i\pv_i}{m^2}\delta(\rv-\rv_i)\delta(\rv'-\rv_i)
  \Big\rangle.
  \label{EQkineticSingular2}
\end{align}
Switching to index notation and denoting Cartesian components by Greek
indices, the $\alpha\gamma$-component of this second-rank tensor is:
\begin{align}
  \sum_{\nu\xi}\nabla_\nu\nabla'_\xi\beta^2
  \Big\langle
  \sum_i \frac{p_i^\alpha p_i^\gamma p_i^\nu p_i^\xi}{m^2}
  \delta(\rv-\rv_i)\delta(\rv'-\rv_i)
  \Big\rangle,
  \label{EQkineticSingular3}
\end{align}
where the sums over the Greek indices run over all Cartesian
components. Carrying out the momentum average over the momentum
tetradic in the integrand requires the following integral over the
Maxwell distribution
\begin{align}
  \frac{\beta^2}{m^2}\int d\pv_i  &
  \frac{
   \e^{-\beta\pv_i^2/(2m)}
  }{(2\pi m/\beta)^{3/2}}
  p_i^\alpha p_i^\gamma p_i^\nu p_i^\xi
  \notag\\
  &=
  \delta_{\alpha\nu}\delta_{\gamma\xi}
  +\delta_{\alpha\xi}\delta_{\gamma\nu}
  +\delta_{\alpha\gamma}\delta_{\nu\xi},
  \label{EQkineticSingular4}
\end{align}
where the combination of the six Kronecker delta symbols on the right
hand side is the isotropic tensor of rank four. The Maxwellian in
\eqr{EQkineticSingular4} is normalized, $\int d\pv_i \e^{-\beta
  \pv_i^2/(2m)}/(2\pi m/\beta)^{3/2}=1$. In the derivation of
\eqr{EQkineticSingular4} one requires both the second moment of each
($\alpha$th) momentum component, $\int d\pv_i (p_i^\alpha)^2\e^{-\beta
  \pv_i^2/(2m)}/(2\pi m/\beta)^{3/2}=m/\beta$, and the fourth moment,
$\int d\pv_i (p_i^\alpha)^4 \e^{-\beta\pv_i^2/(2m)}/(2\pi
m/\beta)^{3/2}=3m^2/\beta^2$.

Recalling the derivative structure of Eq.~\eqref{EQkineticSingular3}
we contract \eqr{EQkineticSingular4} with $\nabla_\nu\nabla_\xi'$,
which yields
\begin{align}
  \sum_{\nu\xi} \nabla_\nu\nabla'_\xi 
  &
   (\delta_{\alpha\nu}\delta_{\gamma\xi}
   +\delta_{\alpha\xi}\delta_{\gamma\nu}
   +\delta_{\alpha\gamma}\delta_{\nu\xi})
   \notag\\
   &=
   (\nabla_\alpha\nabla'_\gamma 
   +\nabla'_\alpha\nabla_\gamma
   + \delta_{\alpha\gamma}\nabla\cdot\nabla')
   \\
   &=
   (\nabla\nabla'+\nabla'\nabla + \unity\nabla\cdot\nabla')_{\alpha\gamma}.
  \label{EQkineticSingular5}
\end{align}
Using this result we can simplify the expression
\eqref{EQkineticSingular3} as
\begin{align}
  (\nabla\nabla'+\nabla'\nabla+\unity\nabla\cdot\nabla')
  \avg{\sum_i\delta(\rv-\rv_i)\delta(\rv'-\rv_i)}.
\end{align}
Hence in summary we obtain the following self contribution:
\begin{align}
   \avg{\nabla \cdot \beta\hat\taub(\rv) & \nabla'\cdot \beta \hat\taub(\rv')}_{\rm self}
   \notag\\
  &= (\nabla\nabla' + \nabla'\nabla +  \unity \nabla\cdot\nabla')
  \rho_2^{\rm self}(\rv,\rv'),
  \label{EQtautauSelf}
\end{align}
where we recall the definition \eqref{EQrho2selfDefinition} of the
two-body self density: $ \rho_2^{\rm self}(\rv,\rv') = \langle\sum_i
\delta(\rv-\rv_i)\delta(\rv'-\rv_i)\rangle$.

We next consider the distinct part of \eqr{EQkineticSingular1}, which
consists of the cases $i \neq j$. In index notation its
$\alpha\gamma$-component is
\begin{align}
  \sum_{\nu\xi}\nabla_\nu \nabla'_\xi \beta^2 \avg{\sum_{ij(\neq)} 
    \frac{p^\alpha_i p^\gamma_j p^\nu_i p^\xi_j}{m^2} 
    \delta(\rv-\rv_i) \delta(\rv'-\rv_j)}.
\label{EQNR1}
\end{align}
As before we carry out the momentum average which requires the
following momentum integrals of particles $i$ and $j$:
\begin{align}
  &\frac{\beta^2}{m^2}
  \int d\pv_i \frac{e^{-\beta\pv_i^2/(2m)}}{(2 \pi m/\beta)^{3/2}}
  p^\alpha_i p^\nu_i \int d\pv_j \frac{e^{-\beta\pv_j^2/(2m)}}
  {(2 \pi m/\beta)^{3/2}} p^\gamma_j  p^\xi_j \nonumber\\
  &\quad= \delta_{\alpha \nu} \delta_{\gamma \xi}. 
\end{align}
Using this result we can re-write \eqr{EQNR1} upon simplifying
according to $ \sum_{\nu\xi} \nabla_\nu \nabla'_\xi
\delta_{\alpha\nu}\delta_{\gamma\xi} = \nabla_\alpha\nabla'_\gamma$ to
hence determine the distinct contribution of the kinetic force
autocorrelator as:
\begin{align}
  &\avg{\nabla \cdot \beta\hat\taub(\rv) 
    \nabla' \cdot \beta\hat\taub(\rv')}_{\text{dist}} \nonumber\\
  &\qquad=\nabla \nabla' \avg{\sum_{ij(\neq)}
    \delta(\rv-\rv_i) \delta(\rv'-\rv_j)}\\
  &\qquad=  \nabla \nabla' \rho_2(\rv,\rv'),
  \label{EQtautauDistinct}
\end{align}
where we have used the definition \eqref{EQrho2distinctDefinition} of
the distinct two-body density.

Hence adding the self contribution \eqref{EQtautauSelf} and the
distinct results \eqref{EQtautauDistinct} we obtain the desired total
kinetic force autocorrelator, multiplied by $\beta^2$, as the
following expression:
\begin{align}
  & \avg{\nabla\cdot\beta\hat\taub(\rv) \nabla'\cdot\beta\hat\taub(\rv')}
  \notag\\&
  = (\nabla\nabla'+\nabla'\nabla+\unity\nabla\cdot\nabla')
  \rho_2^\rmself(\rv,\rv')
  +\nabla\nabla'\rho_2(\rv,\rv'),
  \label{EQtautauTotal}
\end{align}
which we reproduce as \eqr{EQdivTaudivTau} in
Sec.~\ref{SECforceCorrelatorSplitting} of the main text.

\section{Kinetic energy curvature}
\label{SECkineticSelfCurvature}

To second order the momentum transformation
\eqref{EQcanonicalTransformationMomenta} can be written as
\begin{align}
  \pv_i \to \{\unity - \nabla_i \eps(\rv_i) + [\nabla_i\eps(\rv_i)]^2\}
  \cdot \pv_i.
\end{align}
As a consequence, the squared momentum, as is relevant for the kinetic
energy $H_{\rm kin}=\sum_i \pv_i^2/(2m)$, becomes
\begin{align}
  \pv_i\cdot\pv_i 
  \to &
      [\unity - 2\nabla_i\eps_i +
        (\nabla_i\eps_i)^{\sf T}\cdot\nabla_i\eps_i,
       + 2(\nabla_i\eps_i)^2 ]:\pv_i\pv_i,
     \label{EQmomentumSquaredExpansion}
\end{align}
where the colon indicates a double tensor contraction of two matrices
${\sf A}$ and ${\sf B}$, i.e.\ ${\sf A}:{\sf B}= \sum_{\alpha\gamma}
{\sf A}_{\alpha\gamma}{\sf B}_{\gamma\alpha}$, with $\alpha,\gamma$
denoting the Cartesian components; we have used the shorthand
$\eps_i=\eps(\rv_i)$ and as before the superscript $\sf T$
indicates matrix transposition.

We can use the following identities for the linear and quadratic
contribution in the expansion \eqref{EQmomentumSquaredExpansion}. For
the linear contribution, we have
\begin{align}
  [\nabla_i\eps(\rv_i)]^{\sf T} &=
  -\int d\rv \eps(\rv)\nabla\delta(\rv-\rv_i),\\
  \Big(\frac{\delta}{\delta\eps(\rv)}\nabla_i \eps(\rv_i)
  \Big)_{\alpha\gamma\nu}&=
  -\delta_{\alpha\nu}\nabla_\gamma\delta(\rv-\rv_i),
\end{align}
where as above the Greek indices denote the Cartesian components.  We
first consider the following quadratic contribution:
\begin{align}
  [\nabla_i\eps(\rv_i)]^{\sf T}  \cdot\nabla_i\eps(\rv_i) &=
  \notag  \\
  \int  d\rv d\rv'  \eps(\rv)\eps(\rv') &
  \nabla\cdot\nabla'\delta(\rv-\rv_i)\delta(\rv'-\rv_i),
\end{align}
of which the second functional derivative is:
\begin{align}
  \Big(
  \frac{\delta^2}
       {\delta\eps(\rv)\delta\eps(\rv')}\big\{
       [\nabla_i\eps(\rv_i)]^{\sf T}\cdot\nabla_i\eps(\rv_i)\big\}
       \Big)_{\alpha\gamma\nu\xi}&=
       \notag\\
       2\nabla\cdot\nabla'\delta(\rv-\rv_i)&
       \delta(\rv'-\rv_i)\delta_{\alpha\nu}\delta_{\gamma\xi},
       \label{EQdHkinTwo10}
\end{align}
where as before $\nabla'$ indicates the derivative with respect to
$\rv'$.  The further quadratic term is:
\begin{align}
&  [\nabla_i \eps(\rv_i)]^2 =
  \int d\rv d\rv'  \eps(\rv) \eps(\rv') \cdot
  \nabla \nabla' \delta(\rv-\rv_i) \delta(\rv'-\rv_i).
\end{align}
and its second functional derivative with respect to the shifting
field is:
\begin{align}
&  \Big(\frac{\delta^2}{\delta\eps(\rv)\delta\eps(\rv')}
  [\nabla_i\eps(\rv_i)]^2 \Big)_{\alpha\gamma\nu\xi}=
  \notag\\ &\quad
  \big(\delta_{\alpha\xi} \nabla_\gamma \nabla_\nu'
  + \delta_{\gamma\xi} \nabla_\nu \nabla_\alpha' \big)
    \delta(\rv-\rv_i)  \delta(\rv'-\rv_i).
\label{EQdHkinTwo11}
\end{align}

We next build the second functional derivative of the transformed
kinetic energy:
\begin{align}
  &\frac{\delta^2 H_{\rm kin}[\eps]}{\delta\eps(\rv)\delta\eps(\rv')}
  \notag\\&=
  \frac{\delta^2}{\delta\eps(\rv)\delta\eps(\rv')}
  \sum_i [(\nabla_i\eps_i)^{\sf T}\cdot\nabla_i\eps_i
    +2(\nabla_i\eps_i)^2]:
  \frac{\pv_i\pv_i}{2m}\\
  & =\sum_i \Big(\frac{\pv_i\pv_i}{m}\nabla\cdot\nabla'
  + \nabla'\nabla \cdot \frac{\pv_i\pv_i}{m}
  \notag\\&\qquad\qquad
  + \frac{\pv_i\pv_i}{m}\cdot \nabla'\nabla \Big)
  \delta(\rv-\rv_i)\delta(\rv'-\rv_i),
  \label{EQdelta2HkinDeltaEps2Result}
\end{align}
where we have used Eqs.~\eqref{EQdHkinTwo10} and \eqref{EQdHkinTwo11}
in the second step and the factors of 2 have cancelled.  The double
contraction of a fourth- and a second-rank tensor is thereby defined
as $({\sf A}:{\sf B})_{\alpha\gamma}=\sum_{\nu\xi}{\sf
  A}_{\alpha\gamma\nu\xi}{\sf B}_{\xi\nu}$.  Building the thermal
equilibrium average of \eqr{EQdelta2HkinDeltaEps2Result} and
calculating thereby the momentum integrals explicitly yields upon
simplifying the following final compact result:
\begin{align}
  \Big\langle &
  \frac{\delta^2 H_{\rm kin}[\eps]}{\delta\eps(\rv)\delta\eps(\rv')}
  \Big|_{\eps=0}
  \Big\rangle
  &= k_BT(\unity \nabla\cdot\nabla'+2\nabla'\nabla)\rho_2^\rmself(\rv,\rv'),
  \label{EQfunctionalHessianHkinMean}
\end{align}
which is reproduced as \eqr{EQkineticEnergyCurvature} in the main
text.

\section{Potential energy curvature}
\label{SECpotentialEnergyCurvature}

We consider the potential energy contribution to the Hamiltonian
$H_U=u(\rv^N) + \sum_i V_\rmext(\rv_i)$, such that the total
Hamiltonian \eqref{EQHamiltonian2} is $H=H_{\rm kin}+ H_U$. Expressed
in the transformed variables we have
\begin{align}
  H_U[\eps] &= 
  u\big(\rv_1+\eps(\rv_1),\ldots,\rv_N+\eps(\rv_N)\big)
  \notag\\&\quad
  +\sum_i V_\rmext(\rv_i + \eps(\rv_i)).
  \label{EQHUofEps}
\end{align}
The first functional derivative of \eqr{EQHUofEps} is
\begin{align}
  \frac{\delta H_U[\eps]}{\delta\eps(\rv')} &=
  \sum_j\delta(\rv'-\rv_j)
  \nabla_j u\big(\rv_1+\eps(\rv_1),\ldots,\rv_N+\eps(\rv_N)\big)
  \notag\\&\quad
  +\sum_i \delta(\rv'-\rv_i)\nabla_i V_\rmext(\rv_i+\eps(\rv_i)),
  \label{EQdelHIdelEps}
\end{align}
where the Dirac distribution is generated from the functional
derivative $\delta [\rv_j+\eps(\rv_j)]/
\delta\eps(\rv')=\delta(\rv'-\rv_j)\unity$ and the unit matrix then
disapears upon $\unity\cdot\nabla_j=\nabla_j$. The functional
curvature with respect to the shifting field is then obtained as the
functional Hessian (second derivative) of \eqr{EQHUofEps} or
analogously as the first functional derivative of \eqr{EQdelHIdelEps}:
\begin{align}
  \frac{\delta^2 H_U[\eps]}{\delta\eps(\rv)\delta\eps(\rv')}
  \Big|_{\eps=0}
  &=
  \sum_{ij}\delta(\rv-\rv_i)\delta(\rv'-\rv_j)\nabla_i\nabla_j u(\rv^N)
  \notag\\&\quad
  +\sum_i\delta(\rv-\rv_i)\delta(\rv'-\rv_i)\nabla_i\nabla_i V_\rmext(\rv_i),
  \label{EQdelHIdelEpsSecond}
\end{align}
which follows similarly as \eqr{EQdelHIdelEps}. Building then the
equilibrium average of \eqr{EQdelHIdelEpsSecond} yields
\begin{align}
  \avg{\frac{\delta^2 H_U[\eps]}{\delta\eps(\rv)\delta\eps(\rv')}&\Big|_{\eps=0}}
  =
  \notag\\&
  \avg{\sum_{ij}\delta(\rv-\rv_i)\delta(\rv'-\rv_j)\nabla_i\nabla_j u(\rv^N)}
  \notag\\&
  +\rho_2^\rmself(\rv,\rv')\nabla\nabla V_\rmext(\rv),
  \label{EQfunctionalHessianHUMean}
\end{align}
where for the external potential term we have used that here
$\nabla_i\nabla_i=\nabla\nabla$. Together with the corresponding
result \eqref{EQfunctionalHessianHkinMean} for the kinetic energy,
\eqr{EQfunctionalHessianHUMean} forms the functional Hessian of the
full Hamiltonian \eqref{EQHamiltonian2} with respect to the shifting
field. We reproduce the result \eqref{EQfunctionalHessianHUMean} as
\eqr{EQhurra} in the main text.

\section{Sum rules via partial integration}
\label{SECviaPartialIntegration}
As a means of independent verification, we describe an alternative
method to obtain the second-order sum rules, based on more elementary,
but very explicit calculations that do not rely on Noether
invariance. On the downside the thermal symmetry remains hidden.

The argumentation simply rests on suitably chosen integration by parts
in phase space. For a given observable $\hat A(\rv^N,\pv^N)$ the
underlying mechanism is based on two identities. One of them is the
Yvon theorem~\cite{hansen2013, rotenberg2020}:
\begin{align}
  \avg{\nabla_i \hat A} = \avg{\hat A \nabla_i \beta H},
  \label{EQpartialIntegration1}
\end{align}
which can be shown simply by writing out the expression on the left
hand side as $\langle\nabla_i \hat A\rangle=\Tr \Xi^{-1}e^{-\beta
  (H-\mu N)}\nabla_i \hat A = -\Tr \hat A \Xi^{-1} \nabla_i e^{-\beta
  (H-\mu N)}= \Tr \hat A \Xi^{-1} e^{-\beta (H-\mu N)}\nabla_i\beta H
= \langle \hat A \nabla_i\beta H\rangle$, which is the right hand side
of \eqr{EQpartialIntegration1}. In the second step we have assumed
that any boundary terms from the partial integration vanish, because
the external potential contains e.g.\ contributions from impenetrable
container walls.

The second identity involves an additional phase space function $\hat
B$.  We have
\begin{align}
  \avg{\hat A \nabla_i \hat B} =
  \avg{\hat A \hat B \nabla_i \beta H}
  -\avg{\hat B \nabla_i \hat A}.
  \label{EQpartialIntegration2}
\end{align}
The proof of \eqr{EQpartialIntegration2} rests on analogous
argumentation as above, with additionally taking into account the
product rule of differentiation that generates the second term on the
right hand side of Eq.~\eqref{EQpartialIntegration2}. Explicitly, we
start with the left hand side $\langle\hat A \nabla_i \hat
B\rangle=\Tr \Xi^{-1} e^{-\beta (H-\mu N)}\hat A \nabla_i \hat B$ and
integrate by parts, which yields $-\Tr \hat B \Xi^{-1} \nabla_i
e^{-\beta (H-\mu N)}\hat A$. Carrying out the derivative gives two
terms: $\Tr \hat B \Xi^{-1} e^{-\beta (H-\mu N)}(\nabla_i\beta H)\hat
A -\Tr \hat B \Xi^{-1} e^{-\beta (H-\mu N)}\nabla_i\hat A=\langle\hat
A\hat B\nabla_i\beta H\rangle-\langle\hat B\nabla_i\hat A\rangle$,
where we have resorted back to the compact thermal average notation.
Upon ordering of the factors in the first term, we have obtained the
right hand side of \eqr{EQpartialIntegration2}, as desired.

We apply these formal results to the many-body physics under
consideration. We first consider the distinct case. Identifying the
correct starting point is thereby crucial and we take this to be the
double gradient of the (distinct) two-body density
\eqref{EQrho2distinctDefinition}:
\begin{align}
  \nabla\nabla'\rho_2(\rv,\rv') &=
  \nabla\nabla'\avg{{{\sum_{ij(\neq)}}}
   \delta_i\delta_j'
  }.
  \label{EQdoubleDensityGradientDistinct1original}
\end{align}
As before $\nabla'$ indicates the derivative with respect to $\rv'$,
the double sum is denoted using the compact notation
$\sum_{ij(\neq)}=\sum_{i=1}^N\sum_{j=1,j\neq i}^N$ and the delta
  distributions are abbreviated as $\delta_i=\delta(\rv-\rv_i)$ and
$\delta_j'=\delta(\rv'-\rv_j)$. Due to the identity
$\nabla\delta(\rv-\rv_i) = -\nabla_i\delta(\rv-\rv_i)$ we can re-write
the right hand side of \eqr{EQdoubleDensityGradientDistinct1original}
as
\begin{align}
  \avg{\sum_{ij(\neq)}\nabla_i\delta_i\nabla_j\delta_j'}
  &= \avg{\sum_{ij(\neq)}(\nabla_i\beta H)\delta_i\nabla_j\delta_j'}\\
  &= \avg{\sum_{ij(\neq)}(\nabla_i\beta H)(\nabla_j\beta H) \delta_i\delta_j'}
  \notag\\
  &\quad 
  -\avg{{\sum_{ij(\neq)}}(\nabla_i\nabla_j\beta H) \delta_i\delta_j'},
  \label{EQdoubleDensityGradientDistinct1}
\end{align}
where we have integrated by parts in the first step with respect to
$\rv_i$ according to \eqr{EQpartialIntegration1} and in the second
step with respect to $\rv_j$ according to \eqr{EQpartialIntegration2}.

We regroup the double sum inside of the first average in
\eqr{EQdoubleDensityGradientDistinct1} by factorizing $\sum_{ij(\neq)}
(\nabla_i \beta H)(\nabla_j \beta H) \delta_i \delta_j'= \sum_i
\delta_i(\nabla_i\beta H) \sum_{j\neq i}\delta_j'\nabla_j\beta H$.
Taking account of the structure of the Hamiltonian
\eqref{EQHamiltonian2} then allows to identify the expression $-\sum_i
\delta_i \nabla_i H = \hat \Fv_U(\rv)$ as the local potential force
density operator that arises from both interparticle and external
effects, see \eqr{EQFUoperator} for the definition of
$\hat\Fv_U(\rv)$.

We can now express \eqr{EQdoubleDensityGradientDistinct1} succinctly
and hence obtain the distinct part \eqref{EQsumRuleFeFeDistinct} of
the curvature sum rule, which for completeness we supplement by
re-writing the self part \eqref{EQsumRuleFeFeSelf} (to be proven
below):
\begin{align}
  &\avg{\beta\hat\Fv_U(\rv)\beta\hat\Fv_U(\rv')}_{\!\rm dist} =
  \notag\\  &\qquad\qquad 
  \nabla\nabla'\rho_2(\rv,\rv') +
  \avg{{\sum_{ij(\neq)}} (\nabla_i\nabla_j \beta u)\delta_i\delta_j'},
  \label{EQsumRuleFeFeDistinctRepeat}\\
  &\avg{\beta\hat\Fv_U(\rv)\beta\hat\Fv_U(\rv)}_{\!\rm self}
  =
  \notag\\  &\qquad\quad 
  \nabla\nabla\rho(\rv) + \rho(\rv)\nabla\nabla\beta\Vext(\rv)
  +\avg{\sum_i(\nabla_i\nabla_i\beta u)\delta_i}.
  \label{EQsumRuleFeFeSelfRepeat}
\end{align}
The identity \eqref{EQsumRuleFeFeDistinctRepeat} holds for any $\rv,
\rv'$ and \eqr{EQsumRuleFeFeSelfRepeat} is the corresponding self sum
rule, as derived in the following by a proof based on partial phase
space integration, which is similar to the distinct case above.

Analogously to the starting point
\eqref{EQdoubleDensityGradientDistinct1original} for the distinct
correlation identity, we consider the double gradient, but with
respect to the same position $\rv$ taken twice, i.e.\ the Hessian of
the density profile:
\begin{align}
  \nabla\nabla\rho(\rv) &= \nabla\nabla  \avg{\sum_i \delta_i}\\
  &= \avg{ \sum_i \nabla_i\nabla_i \delta_i}\\
  &= \avg{ \sum_i (\nabla_i\beta H)\nabla_i \delta_i}\\
  &= \avg{ \sum_i (\nabla_i\beta H)(\nabla_i\beta H)\delta_i}
\notag\\&\quad
  -\avg{\sum_i (\nabla_i\nabla_i \beta H)\delta_i}.
  \label{EQgradgradDensityDerivation}
\end{align}
In this derivation we have exploited twice that
$\nabla\delta(\rv-\rv_i)=-\nabla_i\delta(\rv-\rv_i)$ and then
integrated by parts first according to \eqr{EQpartialIntegration1} and
then according to \eqr{EQpartialIntegration2}. Identifying
$\hat\Fv_U(\rv)$ in \eqr{EQgradgradDensityDerivation} and re-arranging
the different terms gives \eqr{EQsumRuleFeFeSelfRepeat}.

The present phase space integration route to the sum rules
\eqref{EQsumRuleFeFeDistinctRepeat} and
\eqref{EQsumRuleFeFeSelfRepeat} [or analogously
  Equations~\eqref{EQsumRuleFeFeDistinct} and
  \eqref{EQsumRuleFeFeSelf}] is free of any functional calculus as
required in the derivations in the main text. However, the required
chain of these individual steps is not easy to guess and apparently
these, as we argue, very fundamental results have not been written
down in the existing literature. Hence, while the Noether route comes
at the expense of having to engage with some functional calculus, the
very significant benefit is the simplicity of the starting point,
which here is the second-order invariance
\eqref{EQomegaSecondDerivative1} of the grand potential against
spatially inhomogeneous displacement.

\section{Measuring force-force and force-gradient correlations}
\label{SECappendixMeasuring}

We here provide technical details for the sampling of the force-force
and force-gradient correlations as given in
Eqs.~\eqref{EQforceForceTwoPoint} and
\eqref{EQforceGradientTwoPoint}. Similar to the measurement of the
standard pair correlation function $g(r)$, a loop over particle pairs
is performed in order to calculate an average of a bulk two-body
quantity. However, one not only counts the relative number of
particles separated by a certain distance $r$, but instead also
involves the forces and force gradients that act on the particles in
the calculation of the average.

Let $\rv_i$ and $\rv_j$ be the positions of two particles, such that
$\rv_\parallel = \rv_j-\rv_i$ and $\ev_\parallel =
\rv_\parallel/|\rv_\parallel|$. The calculation of ${\sf g}_\ff(r)$
proceeds straightforwardly by incorporating the forces $\fv_i$ and
$\fv_j$ of particles $i$ and $j$ as given in
\eqr{EQforceForceTwoPoint}. The radial and tangential components are
respectively obtained by the following projections:
\begin{align}
  (\fv_i\fv_j)_\parallel &=
  (\fv_i\cdot\ev_\parallel)(\fv_j\cdot\ev_\parallel),\\
  (\fv_i \fv_j)_\perp &=\frac{
  [\fv_i - (\fv_i \cdot \ev_\parallel) \ev_\parallel] \cdot
  [\fv_j - (\fv_j \cdot \ev_\parallel) \ev_\parallel]}{2},
  \label{EQfifjPerpendicular}
\end{align}
where the factor 1/2 in \eqr{EQfifjPerpendicular} accounts for the two
equal tangential contributions of ${\sf g}_\ff(r)$ in a
three-dimensional bulk fluid (there is only a single radial
component).

For the calculation of ${\sf g}_\gradf(r)$, we employ numerical
differentiation to evaluate the force gradients $\nabla_i \fv_j$ that
appear in \eqr{EQforceGradientTwoPoint}, which is simpler to implement
than analytic Hessians of the interaction potential, in particular for
the more complex models such as the Stillinger-Weber or Gay-Berne
potentials.

As before, a splitting of the force-gradient correlation function
${\sf g}_\gradf(r)$ into its radial and tangential components is
performed. Assuming $\fv_j \nparallel \ev_\parallel$, a tangential
unit vector $\ev_\perp = {\bf t}/|{\bf t}| \perp \ev_\parallel$ can be
constructed by choosing ${\bf t} = \fv_j - (\fv_j \cdot
\ev_\parallel)\ev_\parallel$. The radial and tangential parts of ${\sf
  g}_\gradf(r)$ then follow via
\begin{align}
  (\nabla_i \fv_j)_\parallel &= 
  D_{i,\ev_\parallel}(\fv_j\cdot \ev_\parallel),
  \label{EQmeasuringForceGradientParallel}
  \\
  (\nabla_i \fv_j)_\perp &=
  D_{i, \ev_\perp}(\fv_j \cdot \ev_\perp),
  \label{EQmeasuringForceGradientPerp}
\end{align}
where $D_{i,\ev}$ is a directional derivative operator regarding the
shifting of particle $i$ along the (generic) direction $\ev$.  The
evaluation of Eqs.~\eqref{EQmeasuringForceGradientParallel} and
\eqref{EQmeasuringForceGradientPerp} is performed numerically in the
simulations, i.e., particle $i$ is shifted by a small amount
($\sim10^{-5}\sigma$) to calculate the arising finite difference in
the force $\fv_j$ of particle $j$.

The above procedure applies to isotropic particles. For anisotropic
particles a simple alternative to construct a suitable vector
$\ev_\perp$ can be based on drawing a random unit vector $\ev_{\rm
  ran} \nparallel \ev_\parallel$. Then $\ev_\perp = {\bf t}/|{\bf t}|$
with ${\bf t}=\ev_{\rm ran} - (\ev_{\rm ran}\cdot
\ev_\parallel)\ev_\parallel$, which can be used in
\eqr{EQmeasuringForceGradientPerp}.

\end{document}